\documentclass[man,12pt,noextraspace,floatsintext,natbib]{apa6}
\usepackage{url}
\usepackage{amsfonts}
\usepackage{amsmath,amssymb,color}
\usepackage{xcolor}
\usepackage{graphicx}
\usepackage{enumitem}
\usepackage{subcaption}
\usepackage{caption}
\usepackage{multirow}
\usepackage{bm}
\usepackage{natbib}
\usepackage{MnSymbol}
\usepackage{verbatim}
\usepackage{float}
\usepackage[toc,page]{appendix}
\usepackage[nodisplayskipstretch]{setspace}
\usepackage{adjustbox}
\usepackage{rotating}
\usepackage{diagbox}
\usepackage[linesnumbered,lined,ruled]{algorithm2e}
\usepackage{multicol}
\usepackage{tikz}  
\usetikzlibrary{bayesnet}
\usetikzlibrary{fit,positioning}
\usepackage{arydshln} 
\usepackage{lipsum} 
\usepackage{mwe} %
\usepackage{siunitx}  
\newcommand{\doi}[1]{doi: #1}

\newcommand{\aaa}{{\mathbf r}}

\newcommand{\rrr}{\boldsymbol \gamma}

\newcommand{\ttt}{\boldsymbol \lambda}
\newcommand{\tl}{\lambda}
\newcommand{\tll}{\Lambda}

\newcommand{\ppp}{\boldsymbol \phi}
\newcommand{\xxx}{\boldsymbol \xi}
\newcommand{\hl}{g}
\newcommand{\hhl}{G}
\newcommand{\ar}{r}
\newcommand{\arr}{R}

\newcommand{\bb}{\mbox{$\mathbf b$}}

\newcommand{\ee}{\mbox{$\mathbf e$}}
\newcommand{\ff}{\mbox{$\mathbf f$}}

\newcommand{\zz}{\mathbf z}

\newcommand{\llog}{\mathrm{log}}

\newcommand{\eexp}{\mathrm{exp}}
\newcommand{\ta}{\tilde{a}}
\newcommand{\td}{\tilde{d}}

\newtheorem{theorem}{Theorem}

\title{A Latent Topic Model with Markovian Transition for Process Data}
\author{Haochen Xu \footnote{Affiliation: Fudan University}, Guanhua Fang \footnote{Affiliation: Columbia University, Contact: gf2340@columbia.edu}, Zhiliang Ying \footnote{Affiliation: Columbia University} }
\affiliation{~}
\shorttitle{Latent Topic Analysis}

\begin{document}

\doublespacing

\abstract{
We propose a latent topic model with a Markovian transition for process data, which consist of time-stamped events recorded in a log file. Such data are becoming more widely available in computer-based educational assessment with complex problem solving items. The proposed model can be viewed as an extension of the hierarchical Bayesian topic model with a hidden Markov structure to accommodate the underlying evolution of an examinee's latent state. Using topic transition probabilities along with response times enables us to capture examinees' learning trajectories, making clustering/classification more efficient. A forward-backward variational expectation-maximization (FB-VEM) algorithm is developed to tackle the challenging computational problem. Useful theoretical properties are established under certain asymptotic regimes. The proposed method is applied to a complex problem solving item in 2012 Programme for International Student Assessment (PISA 2012).
}


\maketitle
\setcounter{secnumdepth}{2}


\section{Introduction}

Testing examinees' complex problem-solving (CPS) ability is becoming a prime interest in computer-based assessments.
Recently, a number of prominent large-scale educational assessments include CPS items as their essential components; see, for example, 2012, 2015 and 2018 Programme for International Student Assessment (PISA) \citep{pisa2012, pisa2015}, 2012 Programme for International Assessment of Adult Competencies (PIAAC) \citep{goodman2013literacy}, Assessment and Teaching of 21st Century Skills (ATC21S) \citep{griffin2012assessment}.
The CPS items in these tests are interaction-oriented, requiring students to react to new information adaptively as being received.
A CPS item typically asks examinees to solve a problem in a simulation environment.
In order to arrive at correct answers, examinees may need to learn the environment and acquire knowledge sequentially and interactively. 

For an examinee, the process of solving a CPS item, i.e. the examinee's action sequence, is recorded as a log file.  
These log file data are commonly known as \textit{process data}. Although traditional psychometric models and statistical methods are not directly applicable, there is a growing literature on the process data with varying focuses.
\citet{fischer2011process} reviewed the history of CPS in a variety of research domains and emphasized the importance of information reduction, model building and evaluation in CPS data analysis. 
\citet{halpin2013modelling} proposed to use the Hawkes process to model interactions among examinees in the collaborative CPS items.
\citet{he2015identifying,he2016analyzing} pursued a similar goal by grouping consecutive events into $n$-grams and measuring their association with the outcomes.
\citet{he2017collaborative} discussed the issues and challenges associated with measurement of collaborative problem solving skills by using an example in PISA 2015. 
\citet{polyak2017computational} presented an application of computational psychometrics to collaborative CPS items in the form of continuous Bayesian evidence tracing.
The behavioral paths in an online collaborative problem-solving item are studied by
\citet{vista2017visualising} through event transition graphs.
\citet{xu2018latent} used the latent class model to cluster population based on event histories and response times. 
\citet{qiao2018data} adopted various classification methods to a dataset from PISA 2012 to achieve a better accuracy.
\citet{chen2019statistical} focused on predicting success probability and average residual time for task completion. 

Despite these efforts, modeling and analysis of process data is still in its infancy. Most approaches are ad hoc in nature and there is lack of consensus as to how to develop a comprehensive approach which can handle a large variety of process data. It is desirable to provide a statistical framework that can summarize and handle the important features of process data, i.e., event types and timing (sequence), individual and event type heterogeneity and have parameters with meaningful psychometric interpretation.

In this paper, we propose a hierarchical statistical model with a Markovian structure to characterize both the order/type of events and individual-level effects.
Under this framework, we model the event sequence or process data through a latent Markov chain which represents the evolving latent profiles of the examinee. We assume the first event of the test taker follows some common (baseline) initial distribution. Later events then evolve by following a Markov chain with person-specific transition probabilities and person-specific gap time distributions. We assign a latent topic to each event, with number of topics much smaller than the number of event types, allowing it to have a potential meaning. For computation, a known challenging issue in such modeling, we propose a new method, combining forward-backward algorithm, variational Bayesian method and expectation-maximization (EM) algorithm. Both theoretical and simulation results show that the new method is not only computational tractable, but also provides reasonably good parameter estimation.

The remainder of this paper is organized as follow. In Section \ref{sect::model}, we introduce notation and give model specification. In Section \ref{sec::algorithm}, we present a new forward-backward variational expectation-maximization (FB-VEM) algorithm to obtain parameter estimation. 
In Section \ref{sec:theory}, we establish some theoretical properties for the proposed estimators. 
The simulation results are summarized in Section \ref{sect::sim}. 
In Section \ref{sect::climate}, we apply the proposed method to the process data from a CPS item in PISA 2012. Section \ref{sect::discussion} contains some concluding remarks.


\section{Latent Topic Analysis with Markovian Transition}\label{sect::model}

\subsection{Notation and setting}

Recall that the log file of an examinee contains a sequence of ordered events (actions) coupled with time stamps. We use $N$ to denote the total number of events over testing period $[0, \tau]$, where $\tau$ is the termination time. The observed data sequence for this examinee is denoted by
$\{(e_1, t_1), \ldots, (e_n, t_n), \ldots, (e_N, t_N)\}$
, where $e_n$ is the $n$th event and $t_n$ is the corresponding time stamp. Here $e_n$ takes the value from set $\mathcal E$ which consists of all distinct event types. We use $V$ to denote the cardinality of $\mathcal E$. For notational simplicity, we let $ \bm{t}_{1 : N} = \{t_{n}: n=1,\ldots,N\}$ be the set of ordered event times, where $t_0 = 0 < t_1 < \ldots < t_{N}=\tau$. Also let $\ee_{1 : N} = \{e_n: n = 1, \ldots, N\}$ be the set of the corresponding events. Below, we use the log file of the ``Climate Control'' in PISA 2012 as an example to illustrate the process data structure.

\begin{figure}[H]
\centering
\caption{The Climate Control Item in PISA 2012.}
\includegraphics[width = 0.6\textwidth]{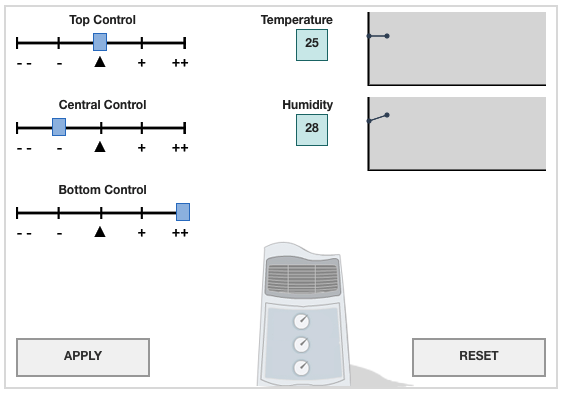}
\label{climate_question}
\end{figure}

\begin{figure}[H]
\centering
\caption{The Climate Control Item Answer Diagram.}
\includegraphics[width = 0.6\textwidth]{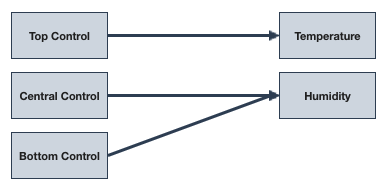}
\label{climate_ans}
\end{figure}

The ``Climate Control'' is a problem solving item from PISA 2012. Around 510,000 15-year-old students from over 60 countries and economies completed the PISA assessment in 2012.  Among them, approximately 85,000 students took the problem solving tests. 
As seen in Figure \ref{climate_question}, the ``Climate Control'' item gives examinees a new air conditioner and asks them to connect three controls to temperature and/or humidity. They can explore the top, central and bottom controls by moving the corresponding sliders and clicking ``APPLY'' or ``RESET'' button. After the click, the temperature and humidity levels are updated in the panel. Once they finish exploring, the examinees need to answer the question, i.e. to draw lines in the diagram (Figure \ref{climate_ans}), connecting sliders to temperature/humidity. 

Table \ref{datastructure} contains the log data of one examinee's event history. This examinee takes 12 actions in 88 seconds. Column ``time'' contains recorded specific times (in second) at which 12 actions were taken. Columns ``top/central/bottom setting'' indicate positions of the three sliders. Detailed explanations of all columns are given in Table 5 in the supplementary.
Because positions of the sliders are updated only at times of ``apply'', we will only consider those events and event times for which the corresponding ``event.type'' is ``apply''. As a result, the set of all distinct events becomes $\mathcal E = \{(0,0,0), \ldots, (2,2,2) \}$, which is the set of all combinations of three slider positions.  
After removing unused rows (1-3,5,7-8,10-12), the cleaned data sequence for this examinee becomes $\{e_1 = (2,0,0), e_2 = (0,2,0), e_3 = (0,2,2)$; $t_1 = 60.1, t_2 = 70.0, t_3 = 80.5\}$. Details about the data cleaning can be found in Section \ref{sect::climate}.

\subsection{Model specification}  

To define our model, we first introduce a (latent) topic sequence, denoted by $\zz_{1:N} = \{z_1, \ldots, z_n, \ldots, z_N\}$. 
We assume $z_{n} \in \mathcal{Z} = \{1, \ldots, K\}$ with $K$ as the number of latent topics.  
In general, we can write 
the density function of the observed data as
\begin{eqnarray} \label{joint-dist}
p(\bm{e}_{1 : N}, \bm{t}_{1 : N})   = \sum_{\zz_{1 : N}} \left[ \prod_{n = 1}^{N} p(e_{n}, t_{n} | \bm{e}_{1 : (n - 1)}, \bm{t}_{1 : (n - 1)}, \zz_{1 : N}) \right] p(\zz_{1 : N}),
\end{eqnarray}
where, for notational simplicity, we let $e_0$ and $t_0$ to denote empty event and time respectively such that $ p(e_1, t_1 |e_0, t_0, \zz_{1:N}) = p(e_1, t_1 | \zz_{1:N}) $. 
Assuming that the $n$th event $(e_n, t_n)$ depends only on the topic transitions from $z_{n - 1}$ to $z_n$ and its preceding time stamp $t_{n - 1}$, we have
\begin{eqnarray}\label{cond-ind}
p(e_{n}, t_{n} | \bm{e}_{1 : (n - 1)}, \bm{t}_{1 : (n - 1)}, \zz_{1 : N})
= p(e_{n}, t_{n} | t_{n - 1}, z_{n - 1}, z_{n}).
\end{eqnarray}
We further assume that $e_n$ and $t_n$ are conditionally independent given $z_{n - 1}$ and $z_n$, the right-hand side of \eqref{cond-ind} becomes
\begin{eqnarray}
p(e_{n}, t_{n} | t_{n - 1}, z_{n - 1}, z_{n}) = p(e_{n}| z_n) p(t_{n} | t_{n - 1}, z_{n - 1}, z_{n}).
\end{eqnarray}
Finally, we assume the latent topic sequence $\{z_n\}_{n = 1}^{N}$ is a Markov chain, i.e. $p(\zz_{1:N}) = \prod_{n = 1}^{N} p(z_n | z_{n - 1}) $, where $p(z_1|z_0)=p(z_1)$. Under these assumptions, (\ref{joint-dist}) becomes
\begin{eqnarray}\label{hmm-dist}
p(\bm{e}_{1 : N}, \bm{t}_{1 : N}) =\sum_{\zz_{1 : N}} 
\left[\prod_{n = 1}^{N} p(e_{n}| z_n) p(t_{n} | t_{n - 1}, z_{n - 1}, z_{n})  \right]
\prod_{n = 1}^{N} p(z_n | z_{n - 1}) .
\end{eqnarray}

We specify the probability distributions on the right hand side of (\ref{hmm-dist}) as follows:
\begin{eqnarray}
e_{n}|z_{n} = k  &\sim& \mathrm{Multinomial}({\bm{b}}_{k}),~ {\bm{b}}_k=(b_{k,1},\ldots,b_{k,V}), \\
z_{n}|z_{n-1} = k^{\prime} &\sim& \mathrm{Multinomial}(\ttt^{k^{\prime}}),~\ttt^{k^{\prime}} = (\tl_{1}^{k^{\prime}},\ldots,\tl_{K}^{k^{\prime}}), \\
z_1 & \sim & \mathrm{Multinomial}(\bm{p}^{0}), \\
t_n - t_{n - 1} | z_{n-1} = k^{\prime}, z_{n} = k, \xi, \hhl & \sim & \mathrm{Exponential}(\xi  e^{\hl_{k^{\prime},k}}) , \label{intensity} \\
\xi | a, d & \sim & \mathrm{Gamma}(a, d).
\end{eqnarray} 
Furthermore, random matrix $\Lambda \equiv (\ttt^{1}, \ldots, \ttt^{K})$ is assumed to follow a Dirichlet prior with parameter $R = (\aaa^{1}, \ldots, \aaa^{K})^{\top}$ such that
\begin{eqnarray}
\ttt^{k^{\prime}} &\sim& \mathrm{Dir}({\aaa}^{k^{\prime}}),~{\aaa}^{k^{\prime}}=(\ar_1^{k^{\prime}},\ldots,\ar_K^{k^{\prime}}). \label{theta_prior}
\end{eqnarray}

In view of  (\ref{hmm-dist}) - (\ref{theta_prior}), we have 
\begin{eqnarray}\label{margin_like}
p(\bm{e}_{1 : N}, \bm{t}_{1 : N}) =& \int_{\tll, \xi} &
\left\{
\sum_{\zz_{1 : N}} 
\prod_{n = 1}^{N} 
p(e_{n}| z_n, {B}) p(t_{n} | t_{n - 1}, z_{n - 1}, z_n, \xi, G) p(z_n | z_{n - 1}, \Lambda ) 
\right\} \nonumber \\
&\cdot& p(\xi|a,d)p(\tll | R) \mathrm{d} \xi  \mathrm{d} \tll
\end{eqnarray}
where $B \equiv (\bm{b}_1, \ldots, \bm{b}_K)$ and $\hhl \equiv (\hl_{k^{\prime},k})_{K \times K}$ are model parameters.

By its definition, ${B}$ is a $K \times V$ matrix that connects the observed event types to latent topics, thereby may be interpreted as ``factor loadings''. It is at population level that does not vary among different examinees. On the other hand, $\tll$ varies with different examinees. Thus for a particular examinee, the corresponding $\Lambda$ may be viewed as a personal transition probability matrix. The intensity function of event time is the product of two components, $H \equiv (e^{\hl_{k^{\prime},k}})_{K \times K}$ and $\xi$. The former, $H$, is at population level, which captures overall examinee's response speed, while the latter, $\xi$, is at individual level, which captures speed heterogeneity among different examinees. In the event history analysis literature \citep{allison1984event, yamaguchi1991event,hougaard1995frailty}, $H$ is interpreted as a fixed effect and $\xi$ is interpreted as a random effect (frailty).  
In our model, a ``topic'' can be viewed as a class of event types sharing with the similar particular meanings. 
Different events, containing various meanings, may belong to distinct topics.
Therefore, topic sequence $\zz_{1:N}$ characterizes the observed event process.     
 
Our model connects the observed data to the latent variables. This is in the spirit of the classical item response theory models \citep[IRT;][]{embretson2013item} and diagnostic classification models \citep[DCMs;][]{templin2010diagnostic}. 
In IRT, examinee's ability is measured by assuming a low-dimensional model structure. The proposed model is also formulated by using the dimension reduction technique.  
In DCM, the $Q$-matrix specifies the relationship between items and latent attributes. In our model, matrix $B$ plays a similar role. It quantifies the relationship between event types and latent topics. 
On the other hand, the proposed model also has its own distinct features. It uses time-stamped event process as the responses, which are no longer binary/multi-categorical. 
For each examinee, the sequence of latent topics can be viewed as his/her latent state. Note that, unlike in IRT/DCM, the length of the sequence is not fixed but depends on the number of actions the examinee takes.

\subsection{Likelihood Function}

By equation \eqref{margin_like}, the likelihood function with $m$ examinees can be written as 
\begin{eqnarray}
&& l(B, \hhl, \bm{p}^0, R, a, d | \{\bm{e}_{1 : N}, \bm{t}_{1 : N}\}_{i=1}^m)
 = 
\prod_{i=1}^m
\bigg\{
\int_{\tll} 
\int_{\xi}
\big\{
\sum_{\zz_{1 : N}} 
\prod_{n = 1}^{N} 
p(e_{n}| z_n, {B})
 \nonumber \\ 
& &
 p(t_{n} | t_{n - 1}, z_{n - 1}, z_n, \xi, G)
p(z_n | z_{n - 1}, \Lambda)  
\big\}p(\xi|a,d)p(\tll | R) \mathrm{d} \xi  \mathrm{d} \tll
\bigg\}. 
\label{all_like}
\end{eqnarray}
In principle, one can get the maximum likelihood estimator (MLE) by maximizing \eqref{all_like}. 
A standard approach is EM algorithm \citep{dempster1977maximum, bailey1994fitting, friedman2001elements}.
However, in practice, it is prohibitively difficult to solve the MLE. 
For this particular case, it is extremely challenging to compute the posterior of latent variables, i.e. 
\begin{eqnarray}\label{true:post}
p(\zz,\tll,\xxx|e,{t},{B},\hhl,\bm{p}^{0}, R, a, d)=\frac{p({t},e,\zz,\tll, \xxx |{B}, \hhl, \bm{p}^{0}, R, a, d)}{p({t}, e|{B}, \hhl, \bm{p}^{0}, R, a, d)}~.
\end{eqnarray}
Specifically, to calculate the denominator of \eqref{true:post}, it requires a large number of summations which grows exponentially fast as the number of events becomes large \citep{blei2003latent}.
 
An alternative approach is variational Bayes  \citep[VB;][]{blei2017variational} method, which is a modern statistical tool to approximate difficult-to-compute probability
densities \citep{blei2003latent, natesan2016bayesian}. In contrast to sampling from true posterior as in the traditional Monte Carlo method, VB postulates a family of distribution, which is assumed to have a much simpler form by reducing many dependency structure, to approximate the true posterior. The estimators are solved by maximizing a different objective function, known as the evidence lower bound (ELBO). 
However, ELBO differs from and is usually smaller than the underlying log-likelihood function. Consequently, the resulting estimation may be biased; for how good ELBO is as a proxy in some special cases, we refer to \citet{hall2011theory}  for the case of Poisson mixture and \citet{you2014variational} for the Bayesian linear model.

In the following two sections, we propose an empirical Bayes-type variational inference with continuous-time hidden Markov processes for event history data. Although there is a literature on variational inference for hidden Markov model \citep{foti2014stochastic, johnson2014stochastic}, the existing work does not cover the current setting in which the events are observed at irregular time points.
In addition, we also establish the usual asymptotic properties, including consistency and normality, of the associated estimators.


\section{Forward-Backward Variational EM Algorithm}\label{sec::algorithm}

In this section, we introduce a forward-backward variational EM (FB-VEM) algorithm to estimate model parameters.
There are two main steps in the proposed algorithm.
\begin{enumerate}
\item For examinee $i$, we consider a variational family $q(\zz_i,\tll_i, \xi_i)$, 
\begin{eqnarray}\label{var-approx}
q(\zz_i,\tll_i, \xi_i)=
q(\xi_i| \ta_i, \td_i)
q(\zz_i|p_i, \kappa_i)
\prod_{k = 1}^{K} q_k(\ttt_i^{k}|\rrr_i^{k}),
\end{eqnarray}
which approximates the conditional joint distribution $p(\zz_i,\tll_i, \xi_i| e,{t},{B}, \hhl, \bm{p}^{0}, R, a, d)$.
Here,
$q(\xi_i| \ta_i, \td_i)$ is the gamma density with shape $\ta_i$ and rate $\td_i$;
$q(\zz_i|p_i, \kappa_i)$ is the (joint) probability density function of vector $\zz_i$ (see \eqref{var:z} in Appendix A); $q_k(\ttt_i^{k}|\rrr_i^{k})$ is a density function of a $K$-dimensional Dirichlet with parameter $\rrr_i^k$. For notational simplicity, we let $q_i = q(\zz_i, \tll_i, \xi_i)$ and $q = \prod_i q_i$ throughout the sequel.
Therefore, $q$ is the probability density function of $(\zz_i, \Lambda_i, \xi_i)_{i=1}^m$ with parameters $(\tilde a_i, \tilde d_i, p_i, \kappa_i, \{\gamma_i^k, k = 1, \ldots, K\})_{i=1}^m$. 
\item  
We define objective function, 
\begin{eqnarray}
EL(q, \eta) \equiv \sum_{i=1}^{m} \big\{\mathbb E_{q_i} \log \big( p(\zz_i, \tll_i, \xi_i, \bm{e}_i, \bm{t}_i| \eta) / q_i \big) \big\},
\end{eqnarray}
where $\eta = (B, \hhl, \bm{p}^0, R, a, d)$.
We maximize $EL(q, \eta)$ with respect to $q$ and $\eta$ by using the coordinate ascent method. We write $EL(q, \eta)$ as $EL$ for simplicity in the remaining of the paper. 
\end{enumerate}

We provide some remarks to end this section.
$EL$ is known as the evidence lower bound (ELBO), which is closely related to Kullback-Leibler (KL) distance \citep{blei2017variational}, i.e.
\begin{eqnarray}\label{elbo:explain}
\sum_i \log p(\zz_i, \tll_i, \xi_i | \bm{e}_i, \bm{t}_i) = EL + \sum_i KL(q_i \|  p(\zz_i, \tll_i, \xi_i| \bm{e}_i, \bm{t}_i, \eta))
\end{eqnarray}
In other words, the log marginal likelihood equals the sum of $EL$ and KL distance between $q$ and true posterior. Therefore, among all distributions in the variational family, a good approximation, $q$, should be close to the true posterior distribution in terms of KL distance.
The computation is similar to that of the EM algorithm, i.e., the model parameters are estimated by solving E-step and M-step alternatively. 
The only difference is that \eqref{elbo:explain} only requires the integration with respect to approximate distribution $q$.
For our choice of variational family, each update has the closed form except for $R$.
Therefore, the computation becomes much simpler as a result.
The complete FB-VEM algorithm is presented in Algorithm \ref{IR},
and the detailed calculation are given in the Appendices A-C.


\section{Theoretical Properties of FB-VEM Algorithm}\label{sec:theory}

In this section, we establish some theoretical results for the FB-VEM algorithm and the parameter estimation. Specifically, we show the convergence to the locally optimal solution of our algorithm in Theorem \ref{thm:algo} and establish the consistency and asymptotic normality of the estimators in Theorems \ref{fix:norm} - \ref{large:norm}.

Recall that the proposed estimator is the maximizer of the following optimization problem 
\begin{eqnarray}\label{est}
& &  (\hat q, \hat {\eta} ) =  \arg \max_{q, \eta} \sum_{i=1}^{m} \big\{\mathbb E_{q_i} \log  \big( p(\zz_i, \tll_i, \xi_i, \bm{e}_i, \bm{t}_i| \eta) / q_i \big) \big\}.
\end{eqnarray}
Since we are only interested in the estimation of ${B}, \hhl$ and $q$, we can assume the priors of $\tll$ and $\xi$ to be fixed without loss of generality. With a slight abuse notation, we let $\eta = ({B}, \hhl)$ be the parameter of interests and $\eta^{\ast} = ({B}^{\ast}, \hhl^{\ast})$ be the true parameter. Furthermore, we assume termination time $\tau$ is the same for all examinees. We denote ELBO by $EL_{\tau}$, which depends on $\tau$ implicitly.

Theorem \ref{thm:algo} gives the local convergence of the FB-VEM algorithm. As a consequence, the proposed estimator will converge to the optimal solution when $EL$ only admits one local maximizer or the starting point is chosen in the neighborhood of the optimum. 
\begin{theorem}\label{thm:algo}
	The FB-VEM algorithm returns a local optimum of \eqref{est}. 
\end{theorem}

The objective function is not the log likelihood but evidence lower bound instead. Evidence lower bound is always smaller than the usual log likelihood. Therefore, we want to know whether or not we can consistently estimate the model parameters including topic-word parameters (i.e. ${B}$) and topic-transition intensity parameters (i.e. $\hhl$); whether or not we can consistently estimate personal transition probability. Our results are stated under two situations: (1) duration $\tau$ is bounded; (2) duration $\tau$ goes to infinity.   
For (1), we show in Theorem \ref{fix:norm} that the estimator will converge, but the limit may be different from the true parameter. For (2), we show that the estimator converges to the true parameter. Furthermore, under certain regularity conditions, the personal-specific transition probabilities can be consistently estimated when $\tau$ goes to infinity. These results are stated in Theorems \ref{large:con} - \ref{large:norm}.

\begin{theorem}\label{fix:norm}
	Under Assumptions A1-A3 given in Appendix D, there exists a consistent estimator $\hat \eta$ such that $\sqrt{m} (\hat \eta - \breve \eta({\tau}) ) \rightsquigarrow N(0, A_1^{-1}(\tau) A_2(\tau) A_1^{-1}(\tau))$. 
\end{theorem}

Theorem \ref{fix:norm} says that the proposed estimator converges to some limit $\breve{\eta}(\tau)$ when time duration $\tau$ is bounded. 
For each fixed $\tau$, it may be viewed as an estimation problem under a mis-specified model, as the estimating equation is constructed via ELBO instead of log likelihood.
As a result, $\breve{\eta}(\tau)$ may be different from true parameter $\eta^{\ast}$, i.e., the estimator is biased when individuals are only observed for a short time. 

However, when individuals are observed for a long time, we can accurately estimate the unobserved personal effect since the measurement and approximation errors will vanish. In that case, we can get consistent estimates of model parameters. The following results hold when both sample size and observation time are large. 

\begin{theorem}\label{large:con}
	Suppose that Assumptions A1-A2, A3'-A4' given in Appendix D hold and that $H_a(\eta)$ admits a unique global maximizer. Then, for any $\delta > 0$, we have that $P(\hat {B} \notin B({B}^{\ast}, \delta)) \rightarrow 0$,  $P(\hat \hhl \notin B(\hhl^{\ast}, \delta)) \rightarrow 0$, $\hat q(\Lambda_i \in B(\Lambda_i^{\ast}, \delta)) \rightarrow_{a.s.} 1$ and $\hat q(\xi_i \in B(\xi_i^{\ast}, \delta)) \rightarrow_{a.s.} 1$ for all $i$ when $m, \tau \rightarrow \infty$. 
\end{theorem}

Theorem \ref{large:con} implies that the evidence lower bound approaches to the log marginal likelihood under a doubly asymptotic regime, i.e. both sample size, $m$, and observation time, $\tau$, is large. Furthermore, we can show that the difference between $EL$ and log marginal likelihood is of order $O(1/\sqrt{\tau})$; see the supplementary. Therefore, we could estimate personal effect and the consistency of topic parameters follows as well.

\begin{theorem}\label{large:norm}
	Under Assumptions A1-A2, A3'-A4' given in the Appendix D and $m = O(\tau^{\delta}) (\delta < 1)$, we have 
	\begin{eqnarray}
	\sqrt m (\hat \eta - \eta^{\ast}) \rightsquigarrow N(0, Q^{-1}) \qquad \textrm{as} ~~ m \rightarrow \infty. 
	\end{eqnarray}
\end{theorem}

One immediate result of Theorem \ref{large:norm} is that the bias of the proposed estimators is of $o(\frac{1}{\sqrt{m}})$, therefore negligible when $m = \tau^{\delta} (\delta < 1)$ and $\tau \rightarrow \infty$. Proofs of Theorems 1-4 are provided in the supplementary.


\section{Simulation Study}\label{sect::sim}

We conducted multiple simulations, three of which are reported here, to assess the performance of the proposed estimators. 
Study 1 emphasizes on the mechanisms of transition structure in the proposed model. Study 2 shows the performance of the proposed method under the classical setting with moderate number of event types. Study 3 evaluates our method under a large-scale setting.
The simulation results show that the proposed method works well and agrees with theoretical findings. 

\subsection{Study 1}
This study considers the situation in which only the sequence of events are used and the time stamps are ignored. It illustrates how the event patterns are captured by the proposed LTA model.

Our set up contains six different event types , ``A'', ``B'', ``C'', ``D'', ``E'' and ``T''. Here, event ``T'' stands for termination, which is always the last event in the process. We assume that ``A'' and ``C'' can only be followed by ``B'' and ``D'' with same probabilities, while all four of them share the same frequency.
We sample six event patterns according to the multinomial distribution shown in Table \ref{eg_1} until the termination event ``T'' is sampled. The corresponding transition probabilities are provided in Table \ref{eg_1_transit}. We generate 100 independent copies of such event processes.
Under this specification, we expect that ``A'' and ``C'' should be in the same topic; ``B'' and ``D'' should be clustered together. This is because that ``A'' and ``B'' are the counterparts of ``C'' and ``D''.

The simulated data is fitted by setting topic number K = 2 and K = 3 respectively. The parameter estimates are given by Tables \ref{result_1} and \ref{result_2}. Here we use norm($\hat{R}$) = $\left( \mathrm{norm}(\hat{r}_{k}^{k^{\prime}}) \right)_{k^{\prime}, k}$ to denote the row-normalized matrix of $\hat{R}$, where $\mathrm{norm}(\hat{r}_{k}^{k^{\prime}}) = \hat{r}_{k}^{k^{\prime}} / \sum_{k = 1}^{K} \hat{r}_{k}^{k^{\prime}}$. From the two tables, we can see that the proposed method perfectly classifies six event types into the topics as expected. Event types ``A'' and ``C'' are in the same topic, while ``B'' and ``D'' are in the other topic. Note that such clustering can not be obtained if we ignore the event transition information.

\subsection{Study 2}
We consider $m=1000$ users, $K=4$ latent topics and $V = 10$ event types in this study. The topic-event matrix $B_{K \times V}$ is constructed in Table \ref{true-beta}, where we highlight the top events in bold font for every topic. We let $a = d = 1$ such that the average random effects of response time is 1.
We set the initial probability of topics to be uniform, i.e. $\bm{p}^{0}=(1/K,\ldots$, $1/K)$.
The parameters $\hhl$ and hyper parameter $R$ are given in Table \ref{true-lambda}. For each user, we simulate the event process according to the the initial probability $\bm{p}^{0}$, the topic-event matrix $B_{K \times V}$ and the intensity parameter $G$ until the $10$th event type occurs. Under this setting, the users would have 500 ($=1/0.002$) events on average in their processes.

We simulate 100 data sets and run 20 times with different initial values for each set.
There are totally $5 \times 10^{5}$ events on average in each data set. \footnote{In this paper, the computation times are reported based on a PC with 2.7 GHz Intel$^\circledR$ Core i5.} Each iteration of the FB-VEM algorithm takes about 3 seconds on average, and the whole estimation procedure completes within 250 iterations. The final estimates $\hat{{B}}$, $\hat{\hhl}$, normalized $\hat{R}$ and their RMSE are given by Tables \ref{estimated-beta} - \ref{estimated-alpha}. From the estimators, we can see that our model successfully captures most of the signals, i.e. the estimated parameters are very close to the truth.  As we have mentioned, the $(k^{\prime}, k)$th entry of the normalized $R$ is the expectation of topic assignment parameter $\tl^{k^{\prime}}_k$. We focus more on the normalized version instead of $\hat{R}$ itself because it gives the probabilities of topic transitions and is more closely related to behavior patterns.

\subsection{Study 3}
In the last study, the performance of our model is evaluated for large data sets. We consider $m=5000$ users, $K = 8$ latent topics and $V = 1000$ event types. We set the $k$th row of ${B}$, $1\leq k \leq K$ as in Table \ref{true-beta-large}. 
We repeat the same procedure as described in Study 2 and get the estimated results. Here each simulated data set contains about $5 \times 10^{5}$ events. It takes around 7.5 seconds to finish one iteration on average, and the whole estimation completes within 300 iterations.


In this study, one way to evaluate the performance of our model is to see whether we can identify the top events with large probabilities and prevent the events with small probabilities from popping up to the top list. Following this idea, we use a cutoff point $b_0$ to divide all events into two groups. Let $c_{k,v}\in \{1,2\}$ denote the true membership of the $v$th event in topic $k$, then we let
\begin{eqnarray}
c_{k,v}=
\begin{cases}
1, b_{k,v}\geq b_0 \\
2, b_{k,v}\leq b_0
\end{cases}.
\end{eqnarray}
The estimated membership $\hat{c}_{k,v}$ is defined in a similar way, equal to 1 if $\hat{{B}}_{k,v}\geq b_0$ and 2 otherwise. We introduce an index
\begin{eqnarray}
CR =\frac{1}{K\cdot V}\sum_{k,v} \mathrm{I}\{c_{k,v}=\hat{c}_{k,v}\}, \label{model-index}
\end{eqnarray}
which takes values from $[0,1]$. This index measures the consistency of the memberships. That is, the larger $CR$ implies the better model fit. 
We let $b_0 = 0.005$, $0.015$, $0.025$, $0.075$, $0.15$ and find that it is more challenging to estimate $c_{k,v}$ when $b_0 = 0.025$. In other words, $CR$ achieves minimum value at $b_0 = 0.025$.
The average of $CR$ for 100 sets of simulation equals $99.89\%$. The result suggests that top events in topics could be successfully detected. 


\section{Application to Climate Control Data}\label{sect::climate}

We apply our method to the ``Climate Control'' item in PISA 2012 as described in Section \ref{sect::model}. The log file of this item contains individual event process history. The data set we use here includes 16920 students, 54.4 \% of whom answered correctly to the item. On average, it takes around 9 actions for a student to explore the item (exclude drawing lines in the diagram), that last for about two minutes.
We remove the ``START\_ITEM'', ``END\_ITEM'' and all ``Diagram'' events. Then we use a 3-dimensional vector to denote the remaining ``apply'' events, with each entry taking a value from \{-2, -1, 0, 1, 2\}. The value here represents the position of the corresponding control slider. For instance, if a student moves the top control to ``2'' while keeping the other two controls at ``$\blacktriangle$'' and then clicks ``APPLY'' button (see the 4th event $e_4$ in Table \ref{datastructure}), then the event is coded as (2, 0, 0).

We fit the model with a series of topic numbers, it turns out that the events with top probabilities are similar across the topics when $K$ is greater than 4. Therefore, the parameter estimates we present here are the results when $K = 4$. The ``Climate Control'' data set contains around $5.3 \times 10^{4}$ events in total. When $K = 4$, each iteration of the FB-VEM algorithm takes about $1$ second on average. It takes less than 600 iterations to finish the whole estimation procedure. To compute the standard errors of estimated parameters, we use parametric bootstrap method by simulating 100 sets of data based on the estimated model. For each set of generated data, we apply the FB-VEM algorithm to obtain the corresponding parameter estimates. We report the standard errors by calculating the standard deviations of 100 sets of estimates.

Table \ref{climate_beta} shows the 4 topics with their top events, and the initial topic distribution is given in Table \ref{climate_theta}. We can see that both Topics 1 and 2 contain event types with at most one moved control at a time, which are the most efficient ways to explore each control. Most examinees will start with events in those two topics according to Table \ref{climate_theta}. Apart from the top events, Topic 3 includes almost all events with more than one moved controls, and the probabilities of them within Topic 3 are quite even. Besides, ``RESET'' seems crucial to this item since it is dominant in Topic 4, though its position in the processes could be different. The estimated $G = ( g_{k^{\prime}, k} )_{k^{\prime}, k}$ in Table \ref{climate_lambda} indicates how fast the examinees would have events from one topic to another. It seems that Topic 2 often comes right after Topic 4, and almost no one would jump to Topic 1 once they have some events from Topic 2. We can also find that it usually takes shorter time to have event types within the same topic.

We further analyze different behavioral patterns of examinees by looking at their person-specific parameters. Here for student, $i$, we use the posterior mean of topic assignment parameter, $\ttt^{k}_i$, as the individual transition probabilities, which could be approximated by the normalized $\rrr^{k}_i$. We denote it as norm($\rrr^{k}_i$). It not only contains information about the topic transition patterns among the whole population, but also captures the personal level variation. We then apply the K-means method to \{ norm($\rrr^{k}_i$), $k = 1, \ldots K$\}. As shown in Table \ref{climate_kmeans}, it turns out that the result is meaningful when the total population is divided into 4 clusters.  According to the average correct rate, the topic transitions do contain significant information about the item. 

We also present the centers of Cluster 1 and Cluster 4 in Table \ref{climate_center} since their average correct rates differ widely. Transition probabilities between Topic 3 and Topic 4 differ substantially across the clusters. 
These two transition matrices also reveal learning trajectories of examinees. Topic 1 (see Table \ref{climate_theta}) is the dominant initial topic. It is mainly about the top control. After the first attempt, around half of the students in Cluster 4 would move on to the rest of the controls and attempt to move multiple bars at the same time (transit from Topic 1 to Topics 2 and 3). They are more likely to keep learning without using ``RESET'' (stay in Topic 2 or 3). Notice that the central and the bottom controls are both about humidity, moving more than one slider at a time could lead to confusion. That might be the reason of their low correct rate.

For students in Cluster 1, after exploring the top control (Topic 1), they tend to either click ``RESET'' (transit from Topic 1 to Topic 4 or stay at Topic 1) and then start to move the second or third control (transit from Topic 1 or 4 to Topic 2), or just go on without clearing up the panel (transit from Topic 1 to Topic 2). Once they reach Topic 2, they could explore the second control, click ``RESET'' to clean up the panel and then try to solve the last control (transit to Topic 4 and then go back to Topic 2). The main strategy behind this systematic behavior path is divide and conquer.


\section{Conclusion}\label{sect::discussion}

In this paper, we propose a latent topic model to analyze process data. Based on a hierarchical Bayesian continuous-time model, we add a hidden Markovian structure. We apply the proposed method to the ``Climate Control'' item in PISA 2012. The proposed model clusters the event types into four latent topics to capture the key features of the test item. Based on the topic transitions of each examinee, we further classify the population into four groups and look into the learning trajectories. It indicates that the strategy known as divide and conquer plays an essential role to solve the item. 

The latent topic model with the proposed FB-VEM algorithm is a general method that could be applied to other CPS items and other kinds of process data such as log files recorded in websites. Once the event type is properly defined, the behavior patterns could be learned through topics and their transitions. Though our approach could be used as a first step to understand the process data, certain domain knowledge is still required to interpret each topic as with most unsupervised methods.

The proposed approach may be extended to include baseline covariates such as gender, nationality and etc.
The latent Markovian structure may also be extended so that the 
current state is related to the entire past history.  
On the computational aspect, since the event processes in the FB-VEM algorithm share only a few common parameters, most user-level parameters could be updated separately in each iteration. Consequently, the distributed algorithm may reduce the computational burden. 
Currently, there is no effective method to compute the standard errors of variational Bayes estimators, which is an important problem for further investigation.

\bibliography{lta}

\newpage

\section{Figures and Tables}

\begin{table}[H]
\centering
\caption{Log Data of an Examinee's Process of Solving the Climate Control Item.}
\label{datastructure}
\begin{adjustbox}{max width=\textwidth}
\begin{tabular}{c lllllllll}
 \hline\hline
event.number & event & time & event.type & top.setting & central.setting & bottom.setting & temp.value & humid.value & diag.state \\ 
  \hline
 1 & START\_ITEM & 0.00 & NULL & NULL         & NULL         & NULL         & NULL         & NULL         & NULL \\ 
 2 & ACER\_EVENT & 40.60 & Diagram & NULL         & NULL         & NULL         & NULL         & NULL         & 000000 \\ 
3 & ACER\_EVENT & 42.60 & Diagram & NULL         & NULL         & NULL         & NULL         & NULL         & 000000 \\ 
4 & ACER\_EVENT & 60.10 & apply & 2            & 0            & 0            & 29           & 25           & NULL \\ 
5 & ACER\_EVENT & 65.00 & Diagram & NULL         & NULL         & NULL         & NULL         & NULL         & 100000 \\ 
6 & ACER\_EVENT & 70.00 & apply & 0            & 2            & 0            & 29           & 27           & NULL \\ 
7 & ACER\_EVENT & 76.40 & Diagram & NULL         & NULL         & NULL         & NULL         & NULL         & 100000 \\ 
8 & ACER\_EVENT & 77.20 & Diagram & NULL         & NULL         & NULL         & NULL         & NULL         & 100100 \\ 
9 & ACER\_EVENT & 80.50 & apply & 0            & 2            & 2            & 29           & 33           & NULL \\ 
10 & ACER\_EVENT & 84.60 & Diagram & NULL         & NULL         & NULL         & NULL         & NULL         & 100100 \\ 
11 & ACER\_EVENT & 85.10 & Diagram & NULL         & NULL         & NULL         & NULL         & NULL         & 100101 \\ 
12 & END\_ITEM & 88.00 & NULL & NULL         & NULL         & NULL         & NULL         & NULL         & NULL \\ 
   \hline
\end{tabular}
\end{adjustbox}
\end{table}

\begin{algorithm}[H]
{
\caption{Forward-Backward Variational EM Algorithm\label{IR}}
\SetKwInOut{Input}{Input} \SetKwInOut{Output}{Output} \SetKwInOut{Initialize}{Initialize}
\setstretch{0.9}
\Input{$t$, $e$.}
\Output{Parameter estimates $\eta=\{{B}, G, \bm{p}^{0}, R, a, d\},\zeta=\{\phi,\tilde{\phi},\gamma, \tilde{\bm{a}}, \tilde{\bm{d}}\}$.}
\Initialize{$p_{i,k}^{k^{\prime}}(i=1:m,k=1:K,k^{\prime}=1:K),\eta$.}
\While{$Q(\eta|\zeta)$ has not converged}{
	\For{$i \in \{1,\ldots,m\}$}{
		\tcp{Update forward and backward probabilities}
		\For{$k \in \{1,\ldots,K\}, n \in \{1,\ldots,N_i\}$}{
			Update $f_{i,n}(k)$ using equation (\ref{forward1}) and (\ref{forward2}) \;
			Update $b_{i,N_i + 1 - n}(k)$ using equation (\ref{backward2}) and (\ref{backward1}) \;
		}
		\tcp{Update the posterior probabilites for latent topics}
		\For{$k \in \{1,\ldots,K\}$, $n \in \{1,\ldots,N_i\}$}{
			Set $\phi_{i,n}^{(k)}$ by equation (\ref{phi-update}) \;
			\For{$l \in \{1,\ldots,K\}$}{
				Set $\tilde{\phi}_{i,n}^{(k^{\prime},k)}$ by equation (\ref{phijoint-update}) \;
			}			
		}
		\tcp{Update variational parameters $\rrr_{i}^{k}$ and transition probabilities}
		\For{$k \in \{1,\ldots,K\}$}{
			Set $\rrr_{i}^{k}$ as in equation (\ref{gamma-update}) \;
			\For{$k^{\prime} \in \{1,\ldots,K\}$}{
				Set $p_{i,k}^{k^{\prime}}$ by equation (\ref{transition-prob}) \;
			}			
		}
		Update $\ta_i, \td_i$ by (\ref{a-update}), (\ref{d-update}) and update $\kappa_i$ \;

	}
	\tcp{We have updated all the parameters in $\zeta=\{\phi,\tilde{\phi},\gamma, \tilde{\bm{a}}, \tilde{\bm{d}}\}$}
	\tcp{Then we apply the EM algorithm}
	\tcp{E-step}
	Get function $Q(\eta|\zeta)$ with updated $\zeta$ defined in equation (\ref{Qfunction}) \;
	\tcp{M-step}
	\For{$k \in \{1,\ldots,K\}$}{
		Set $p^{0}_k$ as (\ref{theta0-update}) \;
		Optimize $Q$ function with respect to $\aaa^{k}$  \;
		\For{$v \in \{1,\ldots,V\}$}{
			Update $b_{k,v}$ by (\ref{beta-update}) \;
		}
		\For{$k^{\prime} \in \{1,\ldots,K\}$}{
			Update $\hl_{k^{\prime},k}$ by (\ref{lambda-update}) \; 
		}

	}
	Compute $Q(\eta|\zeta)$ with updated $\eta=\{{B}, \hhl, \bm{p}^{0}, R, a, d\}$ and $\zeta=\{\phi,\tilde{\phi},\gamma, \tilde{\bm{a}}, \tilde{\bm{d}}\}$ in equation (\ref{Qfunction}) \;
}
}
\end{algorithm}

\begin{table}[H]
\centering
\caption{Study 1: The probabilities of event patterns.} \label{eg_1}
\begin{tabular}{c c c c c c c}
	\hline\hline
	Event Pattern	& AB & AD  & CB  & CD  & E & T \\
	\hline
	Probability   & 8/40 & 8/40 & 8/40 & 8/40 & 7/40 & 1/40 \\
	\hline
\end{tabular}

\caption{Study 1: The transition probabilities of events.} \label{eg_1_transit}
\begin{tabular}{l  llllll}
\hline\hline
  & A   & B   & C   & D   & E     & T \\ 
\hline
A & 0   & 0.5 & 0   & 0.5 & 0     & 0     \\ 
B & 0.4 & 0   & 0.4 & 0   & 0.175 & 0.025 \\ 
C & 0   & 0.5 & 0   & 0.5 & 0     & 0     \\ 
D & 0.4 & 0   & 0.4 & 0   & 0.175 & 0.025 \\ 
E & 0.4 & 0   & 0.4 & 0   & 0.175 & 0.025 \\ 
T & 0.4 & 0   & 0.4 & 0   & 0.175 & 0.025 \\ 
\hline
\end{tabular}

\end{table}

\begin{table}[H]
\centering
\caption{Study 1: Expected and Estimated ${B}$ and $R$ for K = 2.}
\label{result_1}
\begin{tabular}{c ccccccc}
	\hline\hline
	& & A & B & C & D & E & T \\ 
	\hline
	\multirow{2}{*}{$B$} &1 & 0.00 & 0.485 & 0.00 & 0.485 & 0.03 & 0.00 \\ 
								 &2 & 0.41 & 0.00 & 0.41 & 0.00 & 0.15 & 0.03 \\ 
	\hline
	\multirow{2}{*}{$\hat{{B}}$} &1 & 0.00 & 0.485 & 0.00 & 0.485 & 0.03 & 0.00 \\ 
								 &2 & 0.41 & 0.00 & 0.41 & 0.00 & 0.15 & 0.03 \\ 
	\hline
\end{tabular}
\quad
\begin{tabular}{c ccc}
	\hline\hline
								&  & 1    & 2 \\ 
	\hline
	\multirow{2}{*}{norm($R$)}	&1 & 0.00 & 1.00 \\ 
								&2 & 0.87 & 0.13 \\ 
	\hline
	\multirow{2}{*}{norm($\hat{R}$)}	&1 & 0.00 & 1.00 \\ 
										&2 & 0.87 & 0.13 \\ 
	\hline
\end{tabular}
	
\end{table}

\begin{table}[H]
\centering
\caption{Study 1: Expected and Estimated ${B}$ and $R$ for K = 3.}
\label{result_2}
\begin{tabular}{c ccccccc}
	\hline\hline
	& & A & B & C & D & E & T \\ 
	\hline
	\multirow{3}{*}{$B$}        &1 & 0.00 & 0.50 & 0.00 & 0.50 & 0.00 & 0.00 \\ 
					  			&2 & 0.50 & 0.00 & 0.50 & 0.00 & 0.00 & 0.00 \\ 
					  			&3 & 0.00 & 0.00 & 0.00 & 0.00 & 0.87 & 0.13 \\ 
	\hline
	\multirow{3}{*}{$\hat{B}$}  &1 & 0.00 & 0.51 & 0.00 & 0.49 & 0.00 & 0.00 \\ 
						  		&2 & 0.49 & 0.00 & 0.51 & 0.00 & 0.00 & 0.00 \\ 
						  		&3 & 0.00 & 0.00 & 0.00 & 0.00 & 0.87 & 0.13 \\ 
	\hline
\end{tabular}
\quad
\begin{tabular}{c cccc}
	\hline\hline
	& & 1 & 2 & 3 \\ 
	\hline
	\multirow{3}{*}{norm($R$)} 		 &1 & 0.00 & 0.80 & 0.20 \\ 
									 &2 & 1.00 & 0.00 & 0.00 \\ 
									 &3 & 0.00 & 0.80 & 0.20 \\ 
	\hline
	\multirow{3}{*}{norm($\hat{R}$)} &1 & 0.00 & 0.80 & 0.20 \\ 
									 &2 & 1.00 & 0.00 & 0.00 \\ 
									 &3 & 0.00 & 0.80 & 0.20 \\ 
	\hline
\end{tabular}
\end{table}

\begin{table}[H]
\centering
\caption{Study 2: True ${B}$.}
\label{true-beta}
\begin{adjustbox}{max width=\textwidth}
\begin{tabular}{c| c| c c c c c c c c c c}
\hline\hline
 &\diagbox[height=1.2\line]{k~}{~v} &1&2&3&4&5&6&7&8&9&10 \\
 \hline
\multirow{4}{*}{${B}$} &1 & \textbf{0.30} & \textbf{0.30} & \textbf{0.10} & \textbf{0.10} & 0.05 & 0.05 & 0.05 & 0.024 & 0.024 & 0.002 \\
&2 & \textbf{0.10} & \textbf{0.10} & \textbf{0.30} & \textbf{0.30} & 0.05 & 0.05 & 0.05 & 0.024 & 0.024 & 0.002 \\
&3 & \textbf{0.10} & \textbf{0.10} & 0.05 & 0.05 & \textbf{0.30} & \textbf{0.30} & 0.05 & 0.024 & 0.024 & 0.002 \\
&4 & \textbf{0.10} & \textbf{0.10} & 0.05 & 0.05 & 0.05 & 0.024 & \textbf{0.30} & \textbf{0.30} & 0.024 & 0.002 \\
\hline
\end{tabular}
\end{adjustbox}

\end{table}
\begin{table}[H]
\centering
\caption{Study 2: True $\hhl$ and $R$.}
\label{true-lambda}
\begin{adjustbox}{max width=\textwidth}
\begin{tabular}{c| c| r r r r}
\hline\hline 
&\diagbox[height=1.3\line]{$k^{\prime}$~}{~k} & 1 & 2 & 3 & 4 \\
\hline 
\multirow{4}{*}{$\hhl$} &1 & 2 & 1 & -1 & -2 \\
&2 & 1 & 2 & 1 & -1 \\
&3 & -1 & 1 & 2 & 1 \\
&4 & -2 & -1 & 1 & 2 \\
\hline
\end{tabular}
\end{adjustbox}
\quad
\begin{adjustbox}{max width=\textwidth}
\begin{tabular}{c| c| c c c c}
\hline\hline
&\diagbox[height=1.3\line]{$k^{\prime}$~}{~k} & 1 & 2 & 3 & 4 \\
\hline
\multirow{4}{*}{$R$} &1 & 40 & 20 & 5 & 1 \\
&  2 & 1 & 40 & 20 & 5 \\
&  3 & 5 & 1 & 40 & 20 \\
&  4 & 20 & 5 & 1 & 40 \\
\hline
\end{tabular}.
\end{adjustbox}
\end{table}

\begin{table}[H]
\caption{Study 2: Estimated $\hat{{B}}$ and the RMSE ($\times 10^{2}$).}
\label{estimated-beta}
\centering
\begin{adjustbox}{max width=\textwidth}
\begin{tabular}{c | c c c c c c c c c c} 
\hline\hline
\diagbox[height=1.3\line]{k~}{~v} &  1 & 2 &  3 &  4 &  5 &  6 &  7 &  8 &  9 &  10 \\
\hline
1 & \textbf{0.30} & \textbf{0.30} & 0.10 & 0.10 & 0.053 & 0.053 & 0.052 & 0.026 & 0.024 & 0.0021 \\ 
  & (2.1) & (2.2) & (1.2) & (1.2) & (2.1) & (2.0) & (1.5) & (1.6) & (0.089) & (0.047) \\ 
2 & \textbf{0.10} & \textbf{0.10} & \textbf{0.30} & \textbf{0.30} & 0.049 & 0.049 & 0.050 & 0.024 & 0.024 & 0.0020 \\ 
  & (2.1) & (2.1) & (2.0) & (1.9) & (0.27) & (0.26) & (0.11) & (0.12) & (0.065) & (0.025) \\
3 & \textbf{0.10} & \textbf{0.10} & 0.050 & 0.050 & \textbf{0.30} & \textbf{0.30} & 0.049 & 0.023 & 0.024 & 0.0020 \\ 
  & (0.34) & (0.35) & (1.2) & (1.2) & (1.0) & (1.1) & (0.28) & (0.32) & (0.077) & (0.032) \\
4 & \textbf{0.10} & \textbf{0.10} & 0.050 & 0.050 & 0.050 & 0.024 & \textbf{0.30} & \textbf{0.30} & 0.024 & 0.0020 \\ 
  & (0.12) & (0.12) & (0.14) & (0.14) & (0.21) & (0.17) & (0.27) & (0.28) & (0.067) & (0.021) \\
\hline
\end{tabular}
\end{adjustbox}
\end{table}

\begin{table}[H]
\centering
\caption{Simulation Study 2: estimated $\hat{\hhl}$ and the RMSE.}
\label{estimated-lambda}
\begin{adjustbox}{max width=\textwidth}
\begin{tabular}{c|cccc}
\hline \hline
\diagbox[height=1.3\line]{$k^{\prime}$~}{~k} & 1 & 2 & 3 & 4 \\
\hline
1 & 1.9    & 0.97   & -0.93  & -1.9 \\ 
  & (0.59) & (0.34) & (0.37) & (0.56) \\ 
2 & 1.1    & 2.0    & 1.0    & -1.0 \\ 
  & (0.48) & (0.13) & (0.16) & (0.33) \\
3 & -0.96  & 1.0    & 2.0    & 0.99 \\
  & (0.33) & (0.46) & (0.12) & (0.14) \\ 
4 & -1.9   & -1.0   & 0.97   & 2.0 \\ 
  & (0.47) & (0.25) & (0.74) & (0.11) \\
\hline
\end{tabular}
\end{adjustbox}
\end{table}

\begin{table}[H]
\centering
\caption{Simulation Study 2: true $R$, estimated $\hat{R}$ after normalization and the RMSE.}
\label{estimated-alpha}
\begin{tabular}{c|c|cccc}
\hline \hline
 &\diagbox[height=1.3\line]{$k^{\prime}$~}{~k} &1 &2 &3 &4 \\
\hline
\multirow{4}{*}{norm($R$)} & 1 & 0.606 & 0.303 & 0.076 & 0.015 \\ 
&  2 & 0.015 & 0.606 & 0.303 & 0.076 \\ 
&  3 & 0.076 & 0.015 & 0.606 & 0.303 \\ 
&  4 & 0.303 & 0.076 & 0.015 & 0.606 \\ 
\hline
\multirow{8}{*}{norm($\hat{R}$)} 
& 1 & 0.595   & 0.300   & 0.080   & 0.024   \\ 
&	& (0.081) & (0.017) & (0.029) & (0.048) \\
& 2 & 0.023   & 0.608   & 0.294   & 0.075   \\ 
&	& (0.010) & (0.026) & (0.022) & (0.013) \\
& 3 & 0.074   & 0.022   & 0.609   & 0.295   \\ 
&	& (0.014) & (0.014) & (0.016) & (0.017) \\ 
& 4 & 0.289   & 0.083   & 0.016   & 0.612   \\ 
&	& (0.048) & (0.048) & (0.004) & (0.007) \\
\hline
\end{tabular}
\end{table}

\begin{table}[H]
\centering
\caption{Simulation Study 3: True ${B}$.}
\label{true-beta-large}
\begin{adjustbox}{max width=\textwidth}
\begin{tabular}{c| c c c c c c c }
\hline\hline
v &1 &$\cdots$ &$9\times (k-1)$ &$9\times (k-1)+1$
 & $9\times (k-1)+2$ &$9\times (k-1)+3 $ & $9\times (k-1)+4$  \\
\hline
$b_{k,v}$ & $5\times 10^{-4}$ &$\cdots$ & $5\times 10^{-4}$ & 0.3
& 0.1 & 0.05 &0.02  \\
 \hline
v &$9\times (k-1)+5$ &$9\times (k-1)+6$ &$9\times (k-1)+7$ &$9\times (k-1)+8$
 & $\cdots$ &999 & 1000   \\
\hline
$b_{k,v}$ & 0.02 &0.003 &0.001  & $5\times 10^{-4}$
&  $\cdots$ & $5\times 10^{-4}$ &0.01   \\
 \hline
\end{tabular}
\end{adjustbox}
\end{table}

\begin{table}[H]
\centering
\caption{Climate Control item: the 4 topics with their top events.}
\label{climate_beta}
\begin{tabular}{c| c c c c c c c}
\hline\hline
Topic & Top Events   \\
\hline
1  & (1, 0, 0)  & (0, 0, 0) & ``RESET''     \\
\hline
2  & (0, 1, 0)  & (0, 2, 0) & (0, 0, 1) & (0, 0, 2) & (0, 0, 0) &(2, 0, 0)  &(1, 0, 0) \\
\hline
3  & (2, 2, 2)  & (-2, -2, -2) & (1, 1, 1) \\
\hline
4  & ``RESET'' \\
\hline
\end{tabular}
\\~\\
\caption{Climate Control item: estimated $\hhl$ and the standard errors ($\times 10^{2}$).}
\label{climate_lambda}
\begin{tabular}{c| c c c c}
\hline\hline
\diagbox[height = 1.3\line]{$k^{\prime}$~}{~k} & 1 & 2 & 3 & 4 \\ 
\hline
1 & -2.64 & -2.69 & -2.60 & -2.58   \\ 
  & (1.1) & (2.0) & (2.4) & (2.4)   \\
2 & -230.26 & -1.82 & -2.03 & -1.76 \\
  & (1.0) & (0.64) & (1.8) & (0.62)   \\
3 & -2.03 & -2.18 & -1.86 & -2.17   \\ 
  & (7.5) & (3.4) & (0.73) & (2.1)   \\
4 & -3.31 & -1.58 & -2.89 & -2.59   \\
  & (4.7) & (0.67) & (8.7) & (10)  \\
\hline
\end{tabular}

\caption{Climate Control item: estimated $\bm{p}^{0}$ and the standard errors ($\times 10^{3}$).}
\label{climate_theta}
\begin{tabular}{c c c c}
\hline\hline
1 & 2 & 3 & 4 \\
\hline
0.66   & 0.28  & 0.06  & 0.00  \\
(4.7)  & (4.2) & (2.7) & (1.1) \\
\hline
\end{tabular}
\end{table}

\begin{table}[H]
\centering
\caption{Climate Control item: k-means results.}
\label{climate_kmeans}
\begin{tabular}{c | cccc}
\hline\hline
Cluster & 1 & 2 & 3 & 4 \\
\hline
Cluster Size & 4490 & 3706 & 7187 & 1537 \\
Correct Rate  & 81.5 \% & 73.4 \% & 37.0 \% & 11.0 \% \\
\hline
\end{tabular}
\end{table}

\begin{table}[H]
\centering
\caption{Climate Control item: k-means centers of Clusters 1 and 4.}
\label{climate_center}
\begin{tabular}{ccccc}
\hline\hline
Cluster 1 & 1 & 2 & 3 & 4 \\ 
\hline
1 & 0.49 & 0.20 & 0.16 & 0.15 \\ 
2 & 0.00 & 0.20 & 0.05 & 0.75 \\ 
3 & 0.03 & 0.06 & 0.80 & 0.11 \\ 
4 & 0.04 & 0.95 & 0.01 & 0.01 \\ 
\hline
\end{tabular}
\quad
\begin{tabular}{ccccc}
\hline\hline
Cluster 4 & 1 & 2 & 3 & 4 \\ 
\hline
1 & 0.31 & 0.19 & 0.30 & 0.20 \\ 
2 & 0.00 & 0.40 & 0.36 & 0.24 \\ 
3 & 0.32 & 0.08 & 0.35 & 0.25 \\ 
4 & 0.20 & 0.33 & 0.15 & 0.32 \\ 
\hline
\end{tabular}
\end{table}

\newpage

\begin{appendices}

\section{FB-VEM Algorithm}
In this section, we provide a detailed description about the FB-VEM Algorithm. To find a suitable posterior $q(\zz_i,\tll_i, \xi_i)$ which is simple enough and could approximate the true posterior well (equation \ref{true:post}), i.e., 
$$ 
q(\zz_i,\tll_i, \xi_i) \approx p(\zz_i,\tll_i, \xi_i|e,{t},{B}, G, \bm{p}^{0}, R, a, d),
$$ 
we construct posterior $q(\cdot)$ for $\tll_i$, $\zz_i$ and $\xi_i$ separately. 

For $\tll_i$,
we choose a distribution from the following variational family, that is,
\begin{eqnarray}
q(\tll_i)=\prod_{k=1}^{K} q_k(\ttt_i^{k}|\rrr_i^{k}).
\end{eqnarray}
$q_k(\ttt_i^{k}|\rrr_i^{k})$ is set to be a K-dimensional Dirichlet with parameters $\rrr_i^{j}=(\gamma_{i,1}^{k},\ldots,\gamma_{i,K}^{k})$, since the exact conditional distribution of $\ttt_{i}^{k}$ is a Dirichlet
\begin{eqnarray}
p(\ttt_{i}^{k}|z, R)&=&\mathrm{Dir}_K \left(\aaa^{k}+\sum_{n=1}^{N_i}\mathrm{I}\{z_{i,n}=k\}\cdot \zz_{i,n+1} \right), ~~ 1\leq k \leq K.
\end{eqnarray}
Here, $\zz_{i,n+1}$ is a K-vector where the $z_{i,n+1}$th element is 1 and the others all equal to $0$. 
For $\mathbf z_i$, we let it follow a multinomial such that 
\begin{align}\label{var:z}
q(\zz_i|p_i, \kappa_i) \propto 
&  \left( \prod_{n=2}^{N_i} 
\kappa_i e^{\hl_{z_{i,n-1}z_{i,n}}} \exp\{-(t_{i,n} - t_{i,n-1})\kappa_i e^{\hl_{z_{i,n-1}z_{i,n}}}\}
p(e_{i,n}|z_{i,n},{B})p_{i,z_{i,n}}^{z_{i,n}} \right) \nonumber\\
& \cdot  p(e_{i,1}|z_{i,1},{B})p(z_{i,1}|\bm{p}^0).  \nonumber \\
\end{align}
Finally, we set $q(\xi_i | \ta_i, \td_i)$ to be a Gamma distribution with parameter $\ta_i$ and $\td_i$, same as its exact posterior
\begin{eqnarray}
p(\xi_i | t, z, G, a, d) & = & \mathrm{Gamma}(N_i + a - 1, d + \sum_{n = 2}^{N_i} e^{\hl_{z_{i, n - 1}, z_{i, n}}}(t_{i, n} - t_{i, n - 1})).
\end{eqnarray}
After specifying each component, approximate posterior $q(\zz_i, \tll_i, \xi_i)$ has the form
\begin{eqnarray}
q(\zz_i,\tll_i, \xi_i)=
q(\xi_i | \ta_i, \td_i)
q(\zz_i|p_i, \kappa_i)
\prod_{k = 1}^{K} q_k(\ttt_i^{k}|\rrr_i^{k}).
\end{eqnarray}

In order to get the optimal $q(\zz_i, \tll_i, \xi_i)$, we use Kullback-Leibler(KL) divergence as the measure to quantify its distance to $p(\zz_i,\tll_i|\cdot)$, i.e., 
\begin{eqnarray}\label{obj:KL}
\mathrm{KL}(p_i, \rrr_i)=\mathrm{KL}(q(\zz_i,\tll_i)||p(\zz_i,\tll_i|\cdot)).
\end{eqnarray}
The corresponding optimizers are
\begin{eqnarray}
\rrr_i^{k} & = & \aaa^k+\sum_{n=1}^{N_i} (\tilde{\phi}_{i,n}^{(k,1)},\ldots,\tilde{\phi}_{i,n}^{(k,K)})^{\top},  \label{gamma-update} \\
p_{i,k}^{k^{\prime}} & \propto & \eexp \{ \mathbb{E}_{q(\tll_i)} [\log \tll_{i,k}^{k^{\prime}}] \} = \eexp \{ \Psi(\gamma_{i,k}^{k^{\prime}}) -  
\Psi(\sum_{s=1}^{K} \gamma_{i,s}^{k^{\prime}}) \}, \label{transition-prob} \\
\ta_{i} & = & N_i + a - 1,  \label{a-update} \\
\td_{i} & = & d + \sum_{n = 2}^{N_i} (\sum_{k, l} \tilde{\phi}^{(k, l)}_{i, n - 1} e^{\hl_{k, l}}) (t_{i, n} - t_{i, n - 1}), \label{d-update}\\
\kappa_i & = & \tilde a_i/ \tilde d_i. \label{ka-update}
\end{eqnarray}
One of the key elements in the equations above, $\tilde \phi_{i,n}$,  is given in equation \eqref{phijoint-update}.

For the approximate marginal posterior of $z_{i,n}$,  we calculate it by using the forward-backward algorithm.
We let forward and backward functions for subject $i$ at time $t_{i,n}$ be $\ff_{i,n}=(f_{i,n}(1),\ldots,f_{i,n}(K))$ and $\bb_{i,n}=(b_{i,n}(1)$,$\ldots,b_{i,n}(K))$ correspondingly; and let $\ppp_{i,n}=(\phi_{i,n}^{(1)},\ldots,\phi_{i,n}^{(K)})$, $\tilde{\phi}_{i,n}=(\phi_{i,n}^{(k^{\prime},k)})_{k^{\prime},k}$ be the approximate marginal posteriors of $z_{i,n}$ and $(z_{i,n},z_{i,n+1})$ respectively, where
\begin{eqnarray}
\phi_{i,n}^{(k)} &=& p(z_{i,n} = k|{t}_i,\ee_i,{B}, G, \bm{p}^{0}, R), \\
\tilde{\phi}_{i,n}^{(k^{\prime},k)} &=& p(z_{i,n} = k^{\prime},z_{i,n+1}=k|{t}_i,\ee_i,{B}, G, \bm{p}^{0}, R) .
\end{eqnarray}
The posteriors then satisfy
\begin{eqnarray}
\phi_{i,n}^{(k)}&\propto& f_{i,n}(k) \cdot b_{i,n}(k), \label{phi-update}\\
\tilde{\phi}_{i,n}^{(k^{\prime},k)}
&\propto&
f_{i,n}(k^{\prime}) \cdot b_{i,n+1}(k) \cdot p_{i,k}^{k^{\prime}} 
\cdot 
\eexp(\hl_{k^{\prime},k} - \kappa_i e^{\hl_{k^{\prime},k}}(t_{i, n + 1} - t_{i, n}))
\cdot b_{k,e_{i,n+1}}. \label{phijoint-update}
\end{eqnarray}

In the last part of the algorithm, we iteratively do the E-step and the M-step. We let $\eta=\{{B}, G, \bm{p}^{0}, R, a, d\}$, $\zeta=\{\phi,\tilde{\phi}, \gamma, \tilde{\bm{a}}, \tilde{\bm{d}}, p, \kappa\}$. In the E-step we calculate $Q(\eta|\zeta^{(n+1)})$ by
\begin{eqnarray}\label{Qfunction}
\mathbb{E}_{q(z,\theta, \xxx)}[\llog~p({t},e,z,\theta, \xxx)|\zeta^{(n+1)}],
\end{eqnarray}
then in the M-step we solve the following optimization problem
\begin{eqnarray}
\eta^{(n+1)}=\underset{\eta}{\mathrm{arg~max}}~Q(\eta|\zeta^{(n+1)}).
\end{eqnarray}
As two main parts of the FB-VEM algorithm, the forward-backward algorithm and Expectation-Maximization algorithm are described more detailedly in the next two sections.

\section{Forward-Backward Algorithm}\label{sec:f-b-a}

We introduce the forward-backward algorithm in this section and show its application in our model. The algorithm enables us to calculate the posterior distribution of latent variables (states) $\{Y_n\}$ (the latent topic $z_{i,n}$ in our model) given a series of observations $\{X_n\}$ (such as the observed event $e_{i,n}$, the event time $t_{i,n}$) in a hidden Markov model \citep{rabiner1986introduction}. Suppose there are $N$ time stamps in total, we let $Y_{1:n}$ and $X_{1:n}$ denote the latent variables and observations from time $t_{1}$ to $t_{n}$, $1 \leq n \leq N$. A key property of the hidden Markov model is $P(X_{n}|Y_{1:n})=P(X_{n}|Y_{n})$ and $P(X_{n}|Y_{n:N})=P(X_{n}|Y_{n})$, i.e., the observation $X_{n}$ at time $t_{n}$ is independent of other latent variables once given its hidden state $Y_{n}$. Now for a specific $n$, we can apply the property and calculate the conditional probability as
\begin{eqnarray}
P(Y_n|X_{1:N})=P(Y_n|X_{1:n},X_{(n+1):N})\propto P(Y_n|X_{1:n})\cdot P(X_{(n+1):N}|Y_n).
\end{eqnarray}
Here $P(Y_n|X_{1:n})$ and $P(X_{(n+1):N}|Y_n)$ are called forward probability and backward probability, which are two major parts we are trying to obtain in this algorithm. We will derive the recursive formula for these two parts respectively in the following subsections.

\subsection{Forward Probabilities and Forward Functions}
We assume that the latent variable $Y_n$ can take value from $\{1,\ldots,K\}$, then the initial forward probability at time $t_1$ could be calculated by
\begin{eqnarray}
P(Y_1=k|X_{1})\propto P(X_{1},Y_1=k)=P(X_{1}|Y_1=k)\cdot P(Y_1=k), 1\leq k \leq K,
\end{eqnarray}
where $P(Y_1=k)$ only depends on the initial distribution of the latent variables. We introduce the forward functions $\ff_{1}=(f_{1}(1),\ldots,f_{1}(K))$ for simplicity such that
\begin{eqnarray}
f_{1}(k) = P(X_{1}|Y_1=k)\cdot P(Y_1=k).
\end{eqnarray}

At the second time stamp $t_2$, we have
\begin{eqnarray}
P(Y_2=k|X_{1:2}) &\propto& P(X_{1:2},Y_2=k)=\sum_{k^{\prime}=1}^{K}P(X_{1:2},Y_2=k|Y_1= k^{\prime})P(Y_1= k^{\prime}) \nonumber \\
&=& \sum_{k^{\prime}=1}^{K}P(X_{1:2}|Y_2=k,Y_1=k^{\prime})P(Y_2=k|Y_1=k^{\prime})P(Y_1=k^{\prime}). \nonumber
\end{eqnarray}
Notice that
\begin{eqnarray}
P(X_{1:2}|Y_2=k,Y_1= k^{\prime})=P(X_{1}|Y_1= k^{\prime})P(X_{2}|Y_2=k),
\end{eqnarray}
it can be further derived as
\begin{eqnarray}
P(Y_2=k|X_{1:2}) &\propto& \sum_{k^{\prime} =1}^{K}P(X_{2}|Y_2=k)P(Y_2=k|Y_1= k^{\prime})P(X_{1}|Y_1= k^{\prime})P(Y_1= k^{\prime}) \nonumber \\
&=& \sum_{k^{\prime} =1}^{K} f_{1}(k) \cdot p^{k^{\prime}}_{k} \cdot P(X_{2}|Y_2=k)  .
\end{eqnarray}
where $p^{k^{\prime}}_{k}=P(Y_2=k|Y_1= k^{\prime})$ is the transition probability from the hidden state $k^{\prime}$ to $k$. Still we use $\ff_{2}=(f_{2}(1),\ldots,f_{2}(K))$ to denote
\begin{eqnarray}
f_{2}(k) = \sum_{k^{\prime} =1}^{K}f_{1}(k) \cdot p^{k^{\prime}}_{k} \cdot P(X_{2}|Y_2=k).
\end{eqnarray}
In general, given $\ff_{n-1}=(f_{n-1}(1),\ldots,f_{n-1}(K))$ at time $t_{n-1}$, the forward probability at time $t_{n}$ could be obtained as
\begin{eqnarray}
P(Y_n=k|X_{1:n}) &\propto& P(X_{1:n},Y_n=k)=\sum_{k^{\prime} =1}^{K} P(X_{1:n},Y_n=k|Y_{n-1}= k^{\prime})P(Y_{n-1}= k^{\prime}) \nonumber \\
&=& \sum_{k^{\prime} =1}^{K} P(X_{n}|Y_{n}=k)P(Y_{n}=k|Y_{n-1}= k^{\prime})P(X_{1:(n-1)}|Y_{n-1}= k^{\prime})P(Y_{n-1}= k^{\prime}) \nonumber \\
&=& \sum_{k^{\prime} =1}^{K} f_{n-1}(k) \cdot p^{k^{\prime}}_{k} \cdot P(X_{n}|Y_{n}=k)  .
\end{eqnarray}
And the $k$th element of $\ff_{n}=(f_{n}(1),\ldots,f_{n}(K))$ is obtained by
\begin{eqnarray}
f_{n}(k) = \sum_{k^{\prime} =1}^{K} f_{n-1}(k) \cdot p^{k^{\prime}}_{k} \cdot P(X_{n}|Y_{n}=k).
\end{eqnarray}

In our model, the observations for subject $i$ include the detailed events and the response time, so $X_{i,n}=\{e_{i,n},t_{i,n}\}$, while the latent variable is the topic $Y_{i,n}=z_{i,n}$. A main difference between our model and hidden Markov model in this algorithm is that the event time from $t_{i,n-1}$ to $t_{i,n}$, characterized by the intensity function, depends on both the previous and the current topic. It should be regarded as an obervation related to the topic transition. So $P(X_{1:n},Y_n=k|Y_{n-1}= k^{\prime})$ in our case should be
\begin{eqnarray}
&&P(\ee_{i,1:n},\bm{t}_{i,1:n},z_{i,n}=k|z_{i,n-1}= k^{\prime}) \nonumber\\
&=&P(e_{i,n}|z_{i,n}=k)\cdot P(t_{i,n}|t_{i,n-1},z_{i,n}=k,z_{i,n-1}= k^{\prime}) \cdot P(z_{i,n}=k|z_{i,n-1}= k^{\prime}) \nonumber \\
&\cdot& P(\ee_{i,1:(n-1)},\bm{t}_{i,1:(n-1)}|z_{i,n-1}= k^{\prime}).
\end{eqnarray}
Notice that $P(z_{i,1}=k)=p^{0}_{k}$ and $P(e_{i,n}|z_{i,n}=k)=b_{k,e_{i,n}}$. The forward functions of subject i for $n = 1, \ldots, N_i$ are
\begin{eqnarray}
f_{i,1}(k) &=& p^{0}_{k} \cdot b_{k,e_{i,1}}, \label{forward1} \\
f_{i,n}(k) &=& \sum_{k^{\prime} =1}^{K} f_{i,n-1}(k) \cdot p^{k^{\prime}}_{i,k} 
\cdot
\eexp(\lambda_{k^{\prime},k} - \kappa_i e^{\lambda_{k^{\prime},k}}(t_{i, n + 1} - t_{i, n}))
\cdot b_{k,e_{i,n}} . \label{forward2}
\end{eqnarray}

\subsection{Backward Probabilities and Backward Functions}
The backward probabilities start from the last state of the Markov chain (at time $t_N$), then calculate the probability at each time stamp backwards. We assume that the initial backward function $\bb_{N}=(b_{N}(1),\ldots,b_{N}(K))$ is
\begin{eqnarray}
\bb_{N}=(1,\ldots,1).
\end{eqnarray}
This is because there is no more observations after time $t_N$, so we can simply set each of them to be one. At time $t_{N-1}$, by definition the backward probability is
\begin{eqnarray}
P(X_{N}|Y_{N-1}=k)
&=&\sum_{k^{\prime} =1}^{K}P(X_{N}|Y_{N-1}=k,Y_{N}= k^{\prime})\cdot P(Y_{N}= k^{\prime}|Y_{N-1}=k)  \nonumber \\
&=&\sum_{k^{\prime} =1}^{K}b_{N}(k) \cdot p^{k}_{k^{\prime}} \cdot P(X_{N}|Y_{N}= k^{\prime}),
\end{eqnarray}
where $p^{k}_{k^{\prime}}=P(Y_{N}= k^{\prime}|Y_{N-1}=k)$. Then we let each element of the backward function $\bb_{N-1}=(b_{N-1}(1),\ldots,b_{N-1}(K))$ be
\begin{eqnarray}
b_{N-1}(k)= \sum_{k^{\prime} =1}^{K}b_{N}(k) \cdot p^{k}_{k^{\prime}} \cdot P(X_{N}|Y_{N}= k^{\prime}).
\end{eqnarray}

In general, given $\bb_{n+1}=(b_{n+1}(1),\ldots,b_{n+1}(K))$, the backward probability at time $t_{n}$ is
\begin{eqnarray}
P(X_{(n+1):N}|Y_{n}=k)
&=&\sum_{k^{\prime} =1}^{K} P(X_{(n+1):N}|Y_{n}=k,Y_{n+1}= k^{\prime})\cdot P(Y_{n+1}= k^{\prime}|Y_{n}=k)  \nonumber \\
&=&\sum_{k^{\prime} =1}^{K} P(X_{n+1}|Y_{n+1}= k^{\prime}) \cdot
P(X_{(n+2):N}|Y_{n+1}= k^{\prime}) \cdot P(Y_{n+1}= k^{\prime}|Y_{n}=k)  \nonumber \\
&=&\sum_{k^{\prime} =1}^{K} b_{n+1}(k) \cdot p^{k}_{k^{\prime}} \cdot P(X_{n+1}|Y_{n+1}= k^{\prime}),
\end{eqnarray}
so the corresponding backward function is
\begin{eqnarray}
b_{n}(k)=\sum_{k^{\prime} =1}^{K} b_{n+1}(k) \cdot p^{k}_{k^{\prime}} \cdot P(X_{n+1}|Y_{n+1}= k^{\prime}).
\end{eqnarray}
Now we can apply the formulas to our model and get the backward functions from $n = N_i$ to $n = 1$ for each subject $i$ as
\begin{eqnarray}
b_{i,N_i}(k) &=& 1,  \label{backward1} \\
b_{i,n}(k) & = & \sum_{k^{\prime} = 1}^{K} b_{i,n+1}(k^{\prime}) \cdot p_{i,k^{\prime}}^{k} \cdot
\eexp(\lambda_{k, k^{\prime}} - \kappa_i e^{\lambda_{k, k^{\prime}}}(t_{i, n + 1} - t_{i, n}))
\cdot b_{k^{\prime},e_{i,n+1}} .  \label{backward2} 
\end{eqnarray}

\subsection{Posterior Distributions of Latent Variables}
Once we obtain the forward and backward functions, the posterior distribution of each latent variable could be calculated as
\begin{eqnarray}
P(Y_{n}=k|X_{1:N}) \propto P(Y_n|X_{1:n})\cdot P(X_{(n+1):N}|Y_n)\propto f_{n}(k)\cdot b_{n}(k).
\end{eqnarray}
The last thing we need in our algorithm is the joint posterior distribution, which is used to update other parameters in our model. It can be shown that
\begin{eqnarray}
P(Y_{n}=k,Y_{n+1}=l|X_{1:N}) &\propto& P(X_{1:N},Y_{n}=k,Y_{n+1}=l) \nonumber \\
&\propto& P(X_{1:n}|Y_{n}=k) \cdot P(X_{(n+1):N},Y_{n+1}=l|Y_{n}=k) \cdot P(Y_{n}=k) \nonumber \\
&\propto& P(Y_{n}=k|X_{1:n}) \cdot P(X_{(n+2):N}|Y_{n+1}=l) \nonumber \\
&\cdot& P(X_{n+1}|Y_{n+1}=l) \cdot P(Y_{n+1}=l|Y_{n}=k) \nonumber \\
&=& f_{n}(k) \cdot b_{n+1}(l) \cdot p^{k}_{l} \cdot P(X_{n+1}|Y_{n+1}=l)
\end{eqnarray}

Then the corresponding formulas for our model are
\begin{align}
P(z_{i,n}=k|e_{i,1:(N_i)},t_{i,1:(N_i)}) \propto & f_{i,n}(k)\cdot b_{i,n}(k), \\
P(z_{i,n}=k,z_{i,n+1}=l|e_{i,1:(N_i)},t_{i,1:(N_i)}) \propto & f_{i,n}(k)\cdot b_{i,n+1}(l)
\cdot p_{i,l}^{k}  \nonumber \\
&\cdot \eexp(\hl_{k^{\prime},k} - \kappa_i e^{\hl_{k^{\prime},k}}(t_{i, n + 1} - t_{i, n}))
\cdot b_{l,e_{i,n+1}}.
\end{align}


\newpage

\section{Expectation-Maximization Algorithm}\label{sec:em}
In this section, we present the details of parameter estimation using the EM algorithm. There are two steps in the classical EM algorithm, an expectation step (E-step) and a maximization step (M-step). Given the observed data $X$ (such as the observed event $e_{i,n}$, the event time $t_{i,n}$), the unobserved data $Y$ (the latent topic $z_{i,n}$, the topic assignment parameter $\ttt^{j}_{i}$ in our case), and a set of unknown parameters $\eta$ (the topic to event probability matrix ${B}$, intensity-related matrix $\Lambda$, hyper parameter $\alpha$,  etc.), we define the complete-data likelihood as
\begin{equation}
L(\eta;X,Y)=p(X,Y|\eta)~,  \nonumber
\end{equation}
and the log likelihood as $l(\eta;X,Y)=\mathrm{log} L(\eta;X,Y)$.

The EM algorithm iteratively applies the two steps until convergence. In our case, given parameter values $\eta^{(n)}=\{{B}^{(n)}, \hhl^{(n)}, (\bm{p}^{0})^{(n)}, R^{(n)}\}$ obtained in the $n$th iteration, we first update $\zeta=\{\phi,\tilde{\phi},\gamma\}$ using equations (\ref{phi-update}), (\ref{phijoint-update}) and (\ref{gamma-update}) to get $\zeta^{(n+1)}=\{\phi^{(n+1)},\tilde{\phi}^{(n+1)},\gamma^{(n+1)}\}$ in the $(n+1)$th iteration. Then it proceeds as follows:
\begin{enumerate}
	\item (E-step) We calculate the expectation of the log likelihood $l(\ttt;X,Y)$ with respect to the conditional distribution of $Y$ given $X$ and under the current parameter $\zeta^{(n+1)}$,
	\begin{equation}
	Q(\eta|\zeta^{(n+1)})= \mathrm{E}_{Y|X,\zeta^{(n+1)}} l(\eta;X,Y)~. \nonumber
	\end{equation}
	\item (M-step) We find the maximizer of $Q(\eta|\zeta^{(n+1)})$ as a function of $\eta$,
	\begin{equation}
	\eta^{(n+1)}= \underset{\eta}{\mathrm{arg~max}}~Q(\eta|\zeta^{(n+1)})~. \nonumber
	\end{equation}
\end{enumerate}
The explicit form of optimizers in the EM algorithm are given below.

Using the results that
\begin{eqnarray}
\mathbb{E}_{q(\cdot)}[\mathrm{log}~\tl_{i,k}^{k^{\prime}}]&=&\Psi(\gamma^{k^{\prime}}_{i,k})-\Psi(\sum_{s=1}^{K}\gamma^{k^{\prime}}_{i,s}),
\end{eqnarray}
where $\Psi(\cdot)$ is the digamma function, the objective function $Q(\eta|\zeta)$ in the E-step is given by
\begin{eqnarray}
Q(\eta|\zeta)&=&
\mathbb{E}_{q(\cdot)}[\llog~p({t},e,\zz, \tll,\xi)|\zeta]
\nonumber \\
&=& \sum_{i=1}^{m} \left\{
\sum_{n=2}^{N_i} \left[\sum_{k,k^{\prime}=1}^{K} \tilde{\phi}_{i,n-1}^{(k^{\prime},k)}\left(\hl_{k^{\prime},k}
- \frac{\tilde{a}_{i}}{\tilde{d}_{i}} \cdot  e^{\hl_{k^{\prime},k}}(t_{i,n}-t_{i,n-1})+\llog~b_{k,e_{i,n}}\right) \right] \right. \nonumber \\
& + & (N_i + a - 2)[\Psi(\tilde{a}_{i}) - \llog(\tilde{d}_{i})] + a\llog d - \llog \Gamma(a) - d\cdot \frac{\tilde{a}_{i}}{\tilde{d}_{i}} \nonumber \\
&+&
\sum_{n=2}^{N_i}\sum_{k,k^{\prime}=1}^{K} \tilde{\phi}_{i,n-1}^{(k^{\prime},k)} \left(\Psi(\gamma_{i,k}^{k^{\prime}})-\Psi(\sum_{s=1}^{K}\gamma_{i,s}^{k^{\prime}}) \right)
+\sum_{k=1}^{K} \phi_{i,1}^{(k)} \left(\llog~p^{0}_k+ \llog~b_{k,e_{i,1}} \right)
\nonumber \\
&+& \left.
\sum_{k^{\prime}=1}^{K} \sum_{k=1}^{K} \left((\ar_k^{k^{\prime}}-1)(\Psi(\gamma^{k^{\prime}}_{i,k})-\Psi(\sum_{s=1}^{K}\gamma^{k^{\prime}}_{i,s}))\right)
+\sum_{k^{\prime}=1}^{K} \left( \llog~\Gamma(\sum_{k=1}^{K} \ar_{k}^{k^{\prime}})-\sum_{k=1}^{K}\llog~\Gamma(\ar_{k}^{k^{\prime}}) \right) \right\} \nonumber \\
\label{Qfunction}
\end{eqnarray}

In the M-step, we separate terms and maximize with respect to each parameter. The corresponding objective functions are
\begin{eqnarray}
Q({B})&=&\sum_{i=1}^{m}\sum_{n=1}^{N_i} \sum_{k=1}^{K}\sum_{v=1}^{V} \phi_{i,n}^{(k)}\mathrm{I}\{e_{i,n}=v\} \llog~b_{k,v} , \nonumber\\
Q(\arr)&=& \sum_{i=1}^{m} \left\{
\sum_{k^{\prime}=1}^{K} \sum_{k=1}^{K} \left(\ar_k^{k^{\prime}} \cdot (\Psi(\gamma^{k^{\prime}}_{i,k})-\Psi(\sum_{k=1}^{K}\gamma^{k^{\prime}}_{i,k}))\right)
+\sum_{k^{\prime}=1}^{K} \left( \llog~\Gamma(\sum_{k=1}^{K} \ar_{k}^{k^{\prime}})-\sum_{k=1}^{K}\llog~\Gamma(\ar_{k}^{k^{\prime}}) \right)
\right\}, \nonumber \\
Q(G)&=&\sum_{i=1}^{m} \sum_{n=2}^{N_i} \sum_{k,k^{\prime}=1}^{K}\tilde{\phi}_{i,n-1}^{(k^{\prime},k)}\left(\hl_{k^{\prime},k}
- \frac{\tilde{a}_{i}}{\tilde{d}_{i}} \cdot  e^{\lambda_{k^{\prime},k}}(t_{i,n}-t_{i,n-1})\right),  \nonumber\\
Q(\bm{p}^{0})&=&\sum_{i=1}^{m}\sum_{k=1}^{K} \phi_{i,1}^{(k)} \llog~p^{0}_k, \nonumber \\
Q(a) & = & 
\sum_{i=1}^{m}[\Psi(\tilde{a}_{i}) - \llog(\tilde{d}_{i}) + \llog d] \cdot a - m \cdot \llog \Gamma(a) , \nonumber \\
Q(d) & = & m \cdot a \llog d - d\cdot \sum_{i=1}^{m} \frac{\tilde{a}_{i}}{\tilde{d}_{i}}. \nonumber
\end{eqnarray}
The derivatives of $Q({B})$, $Q(\lambda)$ and $Q(\bm{p}^{0})$ are given by
\begin{eqnarray}
\frac{\partial Q({B})}{\partial b_{k,v}}&=&\frac{1}{b_{k,v}}\sum_{i=1}^{m}\sum_{n=1}^{N_i - 1} \phi_{i,n}^{(k)}\mathrm{I}\{e_{i,n}=v\}
-\frac{1}{b_{k,V}}\sum_{i=1}^{m}\sum_{n=1}^{N_i} \phi_{i,n}^{(k)}\mathrm{I}\{e_{i,n}=V\}, \nonumber \\
\frac{\partial Q(\lambda)}{\partial \lambda_{k^{\prime},k}}&=&
\sum_{i=1}^{m} \sum_{n=2}^{N_i} \tilde{\phi}_{i,n-1}^{(k^{\prime},k)} - e^{\lambda_{k^{\prime},k}}\sum_{i=1}^{m} \frac{\tilde{a}_{i}}{\tilde{d}_{i}} \sum_{n=2}^{N_i} \tilde{\phi}_{i,n-1}^{(k^{\prime},k)}(t_{i,n}-t_{i,n-1}),
\nonumber \\
\frac{\partial Q(\bm{p}^{0})}{\partial p^{0}_k}
&=& \sum_{i=1}^{m} \phi_{i,1}^{(k)} \frac{1}{p^{0}_k}
-\sum_{i=1}^{m} \phi_{i,1}^{(K)} \frac{1}{p^{0}_K}, \nonumber \\
\frac{\partial Q(a)}{\partial a}
& = &
\sum_{i = 1}^{m}[\Psi(\tilde{a}_{i}) - \llog(\tilde{d}_{i})] + m \cdot \llog d - m \cdot \Psi(a), \nonumber \\
\frac{\partial Q(d)}{\partial d}
& = &
\frac{m \cdot a}{d} - \sum_{i = 1}^{m} \frac{\tilde{a}_{i}}{\tilde{d}_{i}}. \nonumber
\end{eqnarray}
We set the equations above to be 0, and the corresponding optimizers have closed forms as
\begin{eqnarray}
b_{k,v}&=&\frac{\sum_{i=1}^{m}\sum_{n=1}^{N_i} \phi_{i,n}^{(k)}\mathrm{I}\{e_{i,n}=v\}}{\sum_{i=1}^{m}\sum_{n=1}^{N_i} \phi_{i,n}^{(k)}}, \label{beta-update}\\
\hl_{k^{\prime},k}&=&\llog\frac{\sum_{i=1}^{m} \sum_{n=2}^{N_i} \tilde{\phi}_{i,n-1}^{(k^{\prime},k)}}{\sum_{i=1}^{m} \frac{\tilde{a}_{i}}{\tilde{d}_{i}} \sum_{n=2}^{N_i} \tilde{\phi}_{i,n-1}^{(k^{\prime},k)}(t_{i,n}-t_{i,n-1})}, \label{lambda-update}\\
p^{0}_{k}&=&\frac{\sum_{i=1}^{m} \phi_{i,1}^{(k)}}{m} , \label{theta0-update} \\
d &=& \frac{m \cdot a}{\sum_{i = 1}^{m} \tilde{a}_{i} / \tilde{d}_{i}}. \label{b-update}
\end{eqnarray}
We update $a$ by gradient descent. As for $Q(\arr)$, we calculate its first and second derivatives and use Newton-Raphson algorithm to get the optimizers. The derivatives are
\begin{eqnarray}
\frac{\partial Q(\arr)}{\partial \ar^{k}_{s}}&=&
\sum_{i=1}^{m}\left(\Psi(\gamma^{k}_{i,s})-\Psi(\sum_{l=1}^{K}\gamma^{k}_{i,l})\right)
+m \left(\Psi(\sum_{l=1}^{K}\ar_{l}^{k})-\Psi(\ar_{s}^{k}) \right) \nonumber
\\
\frac{\partial^2 Q(\arr)}{\partial \alpha^{k}_{s} \partial\alpha^{k^{\prime}}_{l}}&=&
m \left( \mathrm{I}\{k=k^{\prime}\} \cdot \Psi^{(1)}(\sum_{r=1}^{K}\ar_{r}^{k})
- \mathrm{I}\{k=k^{\prime},s=l\} \cdot \Psi^{(1)}(\ar_{s}^{k})  \right) \nonumber
\end{eqnarray}
We denote the gradient vectors and Hessian matrces as
\begin{eqnarray}
g_{\alpha}(\aaa^{k})&=&(\frac{\partial Q(\arr)}{\partial \ar^{k}_{s}})_{K\times 1},k=1\ldots K; \nonumber\\
H_{\alpha}(\aaa^{k})&=&(\frac{\partial^2 Q(\arr)}{\partial \ar^{k}_{s} \partial \ar^{k}_{l}})_{K\times K},k=1\ldots K. \nonumber
\end{eqnarray}
In the $(n+1)$th iteration of the Newton-Ralphson method, the estimates are updated as
\begin{eqnarray}
\aaa_{(n+1)}^{k}=\aaa_{(n)}^{k}-H_{\alpha}(\aaa_{(n)}^{k})^{-1}g_{\alpha}(\aaa_{(n)}^{k}). \label{alpha-update} 
\end{eqnarray}
We decompose the matrix $H_{\alpha}(\cdot)$ as
\begin{eqnarray}
H_{\alpha}(\aaa^{k})=m(D(\aaa^{k})+c_k\cdot \bm{1}\times \bm{1}^{\top}), \nonumber
\end{eqnarray}
where
\begin{eqnarray}
D(\aaa^{k})&=&
\mathrm{diag}\{-\Psi^{(1)}(\alpha^{k}_{1}),\ldots,-\Psi^{(1)}(\alpha^{k}_{K})\}, \nonumber \\
c_k&=&\Psi^{(1)}(\sum_{s=1}^{K}\alpha^{k}_{s}). \nonumber
\end{eqnarray}
We can apply the matrix inversion lemma and get
\begin{eqnarray}
m \cdot H_{\alpha}(\aaa^{k})^{-1}&=&
D(\aaa^{k})^{-1}
-\frac{D(\aaa^{k})^{-1}\bm{1}\times \bm{1}^{\top}D(\aaa^{k})^{-1}}
{c_{k}^{-1}+\sum_{s=1}^{K} (d^{k}_{s})^{-1}} \nonumber
\end{eqnarray}
where $d^{k}_{s}$ is the $s$th diagonal element of $D(\aaa^{k})$. We let $g^k_s$ indicate the $s$th element of $g_{\alpha}(\aaa^{k})$ and
\begin{eqnarray}
\tilde{c}_{k}=\frac{\sum_{s=1}^{K}g_s^k/d^{k}_{s}}{c_{k}^{-1}+\sum_{s=1}^{K} (d^{k}_{s})^{-1}},
\nonumber
\end{eqnarray}
now
\begin{eqnarray}
(H_{\alpha}(\aaa^{k})^{-1}g_{\alpha}(\aaa^{k}))_{s}
=\frac{g^k_{s}-\tilde{c}_{k}}{m \cdot d^{k}_{s}}. \nonumber
\end{eqnarray}
We then plug this into the equation \eqref{alpha-update} to get the parameter updates.

\section{Notations and Assumptions}
In this section, we list the notations and assumptions that appear in the main context.
\begin{itemize}
	\item Let $\eta = ({B}, \hhl)$ for notational simplicity and let $\eta^{\ast}$ be the true model parameters. 
	\item Let $y_{n} = (e_{n}, t_{n})$ and $\mathbf Y = (y_1, y_2, \ldots,)$. Let $\mathbf Y_i = (y_{i1}, \ldots, y_{i N_i})$ which is independent copy of $\mathbf Y$.
	\item Under bounded duration setting, it is supposed that $\tau_i \overset{i.i.d}{\sim} f_{\tau}$. $f_{\tau}$ is some density function with bounded support in $\mathrm{R}^{+}$.
	\item We define $EL_{\tau,b}(\eta) = \mathbb E_{\eta^{\ast}} f(Y|\eta)$,
	where the expectation of $Y$ is taken under true parameter $\eta^{\ast} = ({B}^{\ast}, \hhl^{\ast})$. $f(Y|\eta) \equiv \max_{q}\big\{ \mathbb E_q \log p(\mathbf Y|\eta) f_{\tau}(\tau) - \mathbb E_q \log q \big\}$.
	\item Define $\breve{\eta}(\tau)$ to be the $\arg\max_{\eta} EL_{\tau,b}(\eta)$, which represents the best approximate parameter under the proposed variational family.  
	\item Define $A_1(\tau) = \mathbb E_{\eta^{\ast}} (\frac{\partial f(Y|\eta)}{\partial \eta}|_{\breve \eta(\tau)}) (\frac{\partial f(Y|\eta)}{\partial \eta}|_{\breve \eta(\tau)})^T$ and $A_2(\tau) = \mathbb E_{\eta^{\ast}} \frac{\partial^2 f(Y|\eta)}{\partial \eta^2}|_{\breve \eta(\tau)}$.  
	\item Under large duration setting, it is supposed that each each individual has a true underlying personal transition probability $\tll_i^{\ast}$ which defines a aperiodic and irreducible Markov chain and has a true underlying personal frailty $\xi_i^{\ast}$. 
	\item Let $l_{\tau}(\eta, \mathbf Y) = \frac{1}{\tau} \log P(\mathbf Y |\eta)$ and $l_{i, \tau}(\eta, \mathbf Y_i) = \frac{1}{\tau} \log P(\mathbf Y_i |\eta)$.
	\item Let $g_{n}(\eta, \mathbf Y) = \log P(y_{0} | y_{-1}, \ldots y_{-n})$ and $g(\eta, \mathbf Y) = \lim\limits_{n \rightarrow \infty} g_{n}(\eta, \mathbf Y_i)$. 
	Let $g_{i,n}(\eta, \mathbf Y_i) = \log P(y_{i,0} | y_{i,-1}, \ldots y_{i,-n})$ and $g_{i}(\eta, \mathbf Y_i) = \lim\limits_{k \rightarrow \infty} g_{i,n}(\eta, \mathbf Y_i)$, which are the sample versions of $g_{n}(\eta, \mathbf Y)$ and $g(\eta, \mathbf Y)$ respectively. 
	\item Let $s_{\Lambda, \xi} = \lim_{\tau \rightarrow \infty} \frac{N_{\tll, \xi}}{\tau}$, representing the response speed. Let $s_i = \lim_{\tau \rightarrow \infty} \frac{N_i}{\tau_i}$, representing the individual version.
	\item Let $H_{\tll, \xi}(\eta, \xi) = E_{\eta_{\tll, \xi}^{\ast}} g(\eta, \mathbf Y)$. Here, $\eta_{\tll, \xi}^{\ast} = (\tll, \xi, \hhl^{\ast}, {B}^{\ast})$ and the expectation of $\mathbf Y$ is taken under $\eta_{\tll, \xi}^{\ast}$.
	Further, we let $H_a(\eta) = \int s_{\tll, \xi} H_{\tll, \xi}(\eta) p(\tll) p(\xi) d \tll d \xi$. 
\end{itemize} 

Furthermore, we specify the detailed assumptions as followed.  
\begin{itemize}
	\item[A1] (Compactness) Suppose ${B}$ and $\hhl$ lie on a compact parameter space. That is, $b_{k, e} \in [a^{'}, 1-a^{'}]$ and $\hhl_{k, k^{\prime}} \in [a, A]$ for $\forall k, k^{\prime}, e$. 
	\item[A2] The support of $\tll$'s prior distribution is a compact set $\Theta_c \in \{\mathcal S_J\}^{J}$. $\mathcal S_J = \{(\theta_1, \ldots, \theta_J)| \sum_j \theta_j = 1\}$. \\
	The support of $\xi$'s prior distribution is a compact subset of $(0, +\infty)$.
	\item[A3] (Local Identifiability) Both matrices $A_1(\tau)$ and $A_2(\tau)$ are full rank.   
	\item[A3'] (Local Identifiability) $H_{\tll, \xi}(\eta)$ has three time continuous derivatives w.r.t $\eta$ for all $\tll$ and $\xi$.
	Let $Q_1 = \int (\frac{\partial s_{\tll, \xi} H_{\tll, \xi}(\eta)}{\partial \eta}) (\frac{\partial s_{\tll, \xi} H_{\tll, \xi}(\eta)}{\partial \eta})^T p(\tll) p(\xi) d \tll  d \xi$ and $Q_2 = \int  \frac{\partial^2 s_{\tll, \xi} H_{\tll, \xi}(\eta)}{\partial \eta^2} p(\tll) p(\xi) d \tll d \xi$ evaluated at $\eta^{\ast}$. In fact, $Q_1 = Q_2$. We denote both of them as $Q$ which is assumed to be invertible. 
	\item[A4'](Exchangeability) $ \lim_{\tau, m \rightarrow \infty} \frac{1}{m} \sum_i h_{i, \tau}(\eta, \mathbf Y_i) = \lim_{\tau \rightarrow \infty} \lim_{m \rightarrow \infty} \frac{1}{m} \sum_i h_{i, \tau}(\eta, \mathbf Y_i) = \lim_{m \rightarrow \infty} \frac{1}{m} \sum_i  \lim_{\tau \rightarrow \infty} h_{i, \tau}(\eta, \mathbf Y_i)$. $h_{i,\tau}(\eta,\mathbf Y_i)$  could be $l_{i,\tau}(\eta, \mathbf Y_i)$, $\frac{\partial l_{i,\tau}(\eta, \mathbf Y_i)}{\partial \eta}$, $\frac{\partial^2 l_{i,\tau}(\eta, \mathbf Y_i)}{\partial \eta^2}$ or 
	$\frac{\partial^3 l_{i,\tau}(\eta, \mathbf Y_i)}{\partial \eta^3}$.
	\item[A5'] $m \rightarrow \infty$ and $\tau_i = O(m^{r_0})$ for some $r_0 > 1$ for all $i$.
\end{itemize}
The proof of Theorems 1-4 can be found in the supplementary.
\end{appendices}

\end{document}